\documentclass[12pt, aps,amsmath,showpacs,superscriptaddress]{revtex4-1}
\usepackage{graphicx,dcolumn,subfig,caption,bbm,amssymb,amsmath,ulem,indentfirst,amsthm,color}
\usepackage{natbib}
\begin{document}
\title{Collective response in light-matter interactions: The interplay between strong coupling and local dynamics}

\author{Bingyu Cui$^{1,2}$, Abraham Nizan}
\email{anitzan@sas.upenn.edu}
\affiliation{Department of Chemistry, University of Pennsylvania, Philadelphia, Pennsylvania 19104,
USA}
\affiliation{School of Chemistry, Tel Aviv University, Tel Aviv 69978, Israel}
\date{\today}

\begin{abstract}
\noindent A model designed to mimic the implications of the collective optical response of molecular ensembles in optical cavities on molecular vibronic dynamics is investigated. Strong molecule-radiation field coupling is often reached when a large number N of molecules respond collectively to the radiation field. In electronic strong coupling, molecular nuclear dynamics following polariton excitation reflects (a) the timescale separation between the fast electronic and photonic dynamics and the slow nuclear motion on one hand, and (b) the interplay between the collective nature of the molecule-field coupling and the local nature of the molecules nuclear response on the other. The first implies that the electronic excitation takes place, in the spirit of the Born approximation, at an approximately fixed nuclear configuration. The second can be rephrased as the intriguing question, can the collective nature of the optical excitation lead to collective nuclear motion following polariton formation, resulting in so-called polaron decoupled dynamics.  We address this issue by studying the dynamical properties of a simplified Holstein-Tavis-Cummings type model, in which boson modes representing molecular vibrations are replaced by two-level systems while the boson frequency and the vibronic coupling are represented by the coupling between these levels (that induces Rabi oscillations between them) and electronic state dependence of this coupling. We investigate the short-time behavior of this model following polariton excitation as well as its response to CW driving and its density of states spectrum.  We find that, while some aspects of the dynamical behavior appear to adhere to the polaron decoupling picture, the observed dynamics mostly reflect the local nature of the nuclear configuration of the electronic polariton rather than this picture.
\end{abstract}

\maketitle
\section{Introduction}
The manifestation of strong light-matter coupling, either in photochemistry \cite{Assion1998,Barnes2014,Sanvitto2016,Ebbesen2016} or spectroscopy \cite{Ziegler1983,Markel1992,Ziegler1994,Unno1997,Mukamel1995,McHale2017}, has been recently a central theme in chemical research and applications.  Photochemistry, aiming at steering the excited-state dynamics towards a desired product and bypassing unbiased side reactions, is a core technology to develop energy harvesting, converting or solar-energy storing devices \cite{Luk2017}, and strong coupling phenomena have been discussed as possible promising ways towards this end. On the fundamental level, light-matter interaction in a confined dielectric environment on one hand, and the collective nature of the molecular optical response that is often strongly manifested in such environments, present a challenging arena for understanding their role in the energetics and dynamics of molecular processes in such systems \cite{Solano2019,FriskKockum2019,Climent2021,GarciaVidal2021,Nagarajan2021,Xiong2021,Li2022,Dunkelberger2022}.

Common structures used for such setups are Fabry-Perot (FP) cavities where the light-molecular system is confined between two parallel mirrors, plasmonic cavities, where molecules are seated at nanogaps between plasmon sustaining metal particles, and other plasmonic metal structures such as grated metal surfaces or arrays of metal particles. Strong coupling effects in light-molecule interaction in the latter plasmonic structure can be realized at the single molecule level, and may reflect not only the interaction of the molecule with confined optical modes but also changes in molecular electronic structure due to modified Coulomb interactions as well as plasmon induced hot electron generation and electron transfer processes. Here we focus on Fabry-Perot type structures where strong and sometimes ultrastrong coupling phenomena result from the interaction of the electromagnetic field with many ($N = 10^3 – 10^9$) molecules.

The hallmark of strong molecular-field coupling is the observation of Rabi splitting in absorption, emission, scattering, or reflection-transmission of the light, indicating the formation of polaritons - hybrid light-matter states \cite{Guillaume1970,Weisbuch1992,Khitrova1999,Lidzey1999,Lidzey2003,Hobson2002,Schwartz2011,Dintinger2005,Vasa2017,Beane2018,Bisht2018,Melnikau2019}. The collective nature of this interaction is expressed by the linear dependence of this splitting on the square root of the molecular density \cite{Li2022}. Of great interest is the possibility that strong light-matter interaction may affect other molecular dynamic processes, such that energy transfer and other transport phenomena, as well as chemical dynamics. Indeed, many observations of such effects were reported in the last few years, suggesting the possibility to control such processes in suitably constructed optical cavities \cite{Aspelmeyer2014,Li2022,Feist2017,Climent2021,GarciaVidal2021,Nagarajan2021,Xiong2021,Dunkelberger2022,Flick2019,YuenZhou2019,Coccia2020}. Such molecular strong coupling phenomena may be broadly classified as vibronic, where strong coupling originates at resonance between cavity mode(s) and electronic transitions \cite{Ebbesen2016,Solano2019,FriskKockum2019,Climent2021,GarciaVidal2021,Li2022,Mandal2019,Mandal2020,Herrera2016} and purely vibrational, where the cavity resonates with vibrational motions \cite{Ebbesen2016,Thomas2016,Thomas2019, Li2022,Nagarajan2021,Xiong2021,Dunkelberger2022,Hirai2020}.

Theoretical understanding of these observations is at present incomplete. Observed chemical consequences of vibrational strong coupling in the electronic ground state and in the absence of incident light still await theoretical descriptions despite recent efforts \cite{delPino2015,delPino2016,KnaCohen2019,Flick2018,FRibeiro2018,Xiang2018,CamposGonzalezAngulo2019,Climent2020,Li2020,Li2021a,Li2021,Li2021c,Yang2021,cao2022,Mandal2022}. More progress has been made in the theoretical interpretation of photochemical processes initiated by molecular electronic excitations. However, while electronic strong coupling effects on the dynamics of one or two molecules are believed to be fairly well understood \cite{Galego2015,Chikkaraddy2016,Li2022}, actual observations are made on many-molecule systems in which strong coupling arises from a collective molecular response. Similar collective behaviors in subsequent energy transfer phenomena have been discussed \cite{GmezCastao2019}, and a related theoretical activity has focused on possible applications in energy storage and release in so called quantum batteries \cite{Campaioli2018,Bhattacharjee2021}, however these studies usually focus on two level models of the excited species, and a full understanding of the implications and consequences of collective optical driving of molecular systems is still lacking. At its core, the issue concerns the apparent disparity between the collective nature of the molecular optical response, in particular the collective nature of molecular polariton excitation, which is prominently expressed by the observed density dependence of the Rabi splitting, and the local nature of subsequent molecular processes including chemical reactions. In other words, a central question is if and how the internal dynamics of a single molecule is affected by the preceding excitation and formation of a collective molecular (or hybrid molecular-optical) state. To address this question one needs to go beyond the Tavis-Cummings (TC) model of $N$ 2-level atoms interacting with an electromagnetic field mode \cite{Tavis1968,Tavis1969} that is often used to discuss polariton formation. The Holstein-Tavis-Cummings (HTC),
\begin{equation}
    \hat{H}_{HTC}=\hbar\omega_c\hat{a}^\dagger\hat{a}+\hbar\sum_{j=1}^N\left[\omega_{xg}\hat{\sigma}^+_j\hat{\sigma}^-_j+\frac{g}{2}(\hat{a}^\dagger\hat{\sigma}^-_j+\hat{\sigma}^+_j\hat{a})+\omega_v\hat{b}^\dagger_j\hat{b}_j+\lambda\hat{\sigma}^+_j\hat{\sigma}^-_j(\hat{b}^\dagger_j+\hat{b}_j)\right]+\sum_{j=1,i\neq j}^N J_{ij}\hat{\sigma}^+_i\hat{\sigma}^-_j
    \label{eq:HHTC}
\end{equation}
which extends the Holstein polaron model \cite{Holstein1959,Holstein1959b} to include coupling with a cavity mode, has been used to this end \cite{Galego2016,Herrera2016,Galego2015,Wu2016,Spano2015,Zeb2017}. In Eq. \eqref{eq:HHTC}, each of the molecules is a two-electronic-state entity, and in its minimal version, a single exciton case is considered. All molecules are taken identical and the disorder associated with the distribution of molecular orientations relative to the cavity modes is also ignored. In Eq. \eqref{eq:HHTC}, the operator $\hat{a}$ ($\hat{a}^\dagger$) annihilates (creates) a photon of a cavity mode of frequency $\omega_c$ while $\hat{\sigma}_j=|g_j\rangle\langle e_j|$ and $\hat{\sigma}^\dagger_j=|e_j\rangle\langle g_j|$ respectively affect the upward and downward transitions between the lower $|g_j\rangle$ and upper $|e_j\rangle$ electronic states of molecule $j$, $\hbar\omega_{xg}$ is the molecular electronic transition energy and $g$ is the molecule-cavity mode interaction matrix element, taken to be the same for all molecules. The above are parameters of the TC Hamiltonian. In addition, each molecule is associated with one harmonic oscillator of frequency $\omega_v$, described (for molecule $j$) by the raising and lowering operators $\hat{b}_j^\dagger$ $(\hat{b}_j)$ and vibronic interaction given by the parameter $\lambda$, which corresponds to a shift of the molecular nuclear harmonic surfaces between the ground and excited electronic states. The last term in Eq. \eqref{eq:HHTC}, an intermolecular interaction that leads to excitation energy transfer between molecules, is ignored in the TC model and, as done in several recent works \cite{Spano2015,Herrera2016}, will be ignored in the calculation described below \footnote{\label{ft1} It should be kept in mind that in molecular crystals as well as ordered dye aggregates intermolecular interactions lead to collective excitonic response \cite{Spano2011} which may be modified by a cavity environment. The present analysis disregards such interaction in order to focus on the cavity effect}, in order to focus on the effect of coupling to the cavity mode.

In this model, outside the cavity ($g=0$), starting from the ground vibrational state on the lower electronic surface, a sudden broad band excitation projects the wavefunction vertically onto the upper potential surface and a sudden change of force is experienced by the internal motion of the excited molecule. Inside the cavity, if the collective Rabi splitting, $\Omega_R=g\sqrt{N}$, is large relative to the width of the Frank Condon envelope, we expect to see this line split into an upper and lower polariton transitions. The implication for the nuclear dynamics following such excitations has been discussed by several workers \cite{Spano2015,Herrera2016,Zeb2017,Feist2015,Galego2015,Luk2017,Feist2017,Flick2017,Rubio2018,Flick2019,Haugland2020,Haugland2021,Mandal2019,Mandal2020}, leading to interesting observations about the possible implication of strong coupling to a cavity mode on the standard Born-Oppenheimer (BO) picture of molecular nuclear dynamics. The focus here is on electronic strong coupling where, under the standard separation of timescales to electronic (fast) and nuclear (slow) molecular motions, the cavity mode(s) belongs to the fast group together with the molecular electronic dynamics. This has led Spano \cite{Spano2015}, and later Herrera and Spano \cite{Herrera2016} to suggest that in the large molecular number $(N)$ limit, the nuclear potential surface associated with $a$-polariton state is similar (up to deviations of order $N^{-1}$) to the ground state potential surface, with strong implications for subsequent interstate electronic transitions including electron transfer processes. The argument is similar to that used for processes involving excitons formed in molecular J and H aggregates. Such aggregates are characterized by strong intermolecular dipolar interactions that lead to strong delocalization of the molecular excitation and to electronic states that are characterized by their excitonic quasi-momentum $k$ rather than as locally excited molecules. Under such conditions the bright, $k=0$, exciton mode is energetically separated from other modes, making the corresponding excitonic Born-Oppenheimer surface relevant for analyzing subsequent electronic processes. Because a single exciton state involving $N$ molecules is associated with $N-1$ ground state molecules, its potential surface indeed mimics the ground state molecular surface up to $N^{-1}$ corrections, implying a substantially smaller effect of electronic-vibrational coupling (an effective polaron decoupling, otherwise referred to as vibronic decoupling) on molecular dynamic processes following polariton excitation. Herrera and Spano \cite{Herrera2016} have asserted that when the coupling of the molecular system to the cavity is strong enough, $g\gg\omega_v$, the Born-Oppenheimer approximation can be invoked to predict a similar behavior. Later studies \cite{Luk2017,Dunkelberger2016,Galego2015,Herrera2017} have indeed confirmed the existence of polaron decoupling, however in a restrictive form – provided that the Rabi splitting is large relative to the reorganization energy associated with the vibronic coupling. It has been suggested that such polaron decoupling may have dramatic consequences for photophysical and photochemical processes initiated by polariton excitation \cite{Galego2016,Herrera2016,Galego2017,Mauro2021,Herrera2017,Mandal2019,Avramenko2020}.

It is easy to associate such arguments as described above with a single or a few molecules that are strongly coupled to a cavity mode \cite{Feist2017,Feist2015} since conceptually there is no difference between the standard molecular Born-Oppenheimer approximation and its cavity counterpart: when the characteristic cavity timescale $\omega_c^{-1}$ is of the order of characteristic electronic motions, both cavity and electronic dynamics act as a quickly adjusting background for the slow nuclear motions, creating modified potential energy surfaces for the latter. Crossing of these modified surfaces \cite{Fbri2021} can be handled as in standard non-adiabatic calculations, e.g. using surface hopping techniques, does providing a cavity-Born-Oppenheimer framework for calculating and predicting possible cavity effects on molecular dynamics, including chemical processes \cite{Flick2017,Flick2017b,Flick2018,Lacombe2019,Flick2020,Dunkelberger2016,Semenov2019}. However, in FP cavities, the coupling of the electromagnetic field to a single or a few molecules is usually not strong enough to realize such effects, and strong coupling is usually achieved by the collective coupling of a cavity mode to a large number of molecules \footnote{Strong coupling involving a small number of molecules can be realized in plasmonic cavities, yet cavity effects in such structures may arise also from several other dynamical processes, including modifications of electrostatic interactions by the cavity environment, plasmon-driven electron tunneling and plasmon-induced formation of hot electrons.}. In this case, the electronic dynamics is dominated by two different timescales. The timescale associated with the Rabi splitting, $\Omega_R=g\sqrt{N}$ ($g$, the molecule-cavity mode coupling in Eq. \eqref{eq:HHTC}, is the Rabi splitting associated with a single molecule) can be of order $\tau_R=\Omega_R^{-1}\sim10$fs or even shorter \cite{Li2022}, much smaller than characteristic molecular nuclear time $\tau_{Nuc}$. However, the timescale associated with energy transfer between molecules is of order $\tau_{ET}=\tau_R\sqrt{N}=g^{-1}$ (see Sec. 1 in Supplementary Information (SI)) where $N$  can be of order $10^5$ as a conservative estimate. These timescales thus satisfy 
\begin{equation}
    \tau_R<\tau_{Nuc}<\tau_{ET}.
    \label{eq:timediff}
\end{equation}
For this reason, the implications of the timescale separation between the cavity-coupled molecular electronic dynamics and the molecular nuclear dynamics should be scrutinized more carefully. For the Born-Oppenheimer approximation to be valid in such circumstances, the slow timescale $\tau_{ET}$ should not be expressed in the dynamics that determine the nuclear potential surface. On the other hand, local processes that are represented in the Hamiltonian \eqref{eq:HHTC} by the last two terms in the mid-bracket (oscillators local to the individual molecules that respond to the excitation state of “their” molecules) tend to localize the excitation on individual molecules. The interplay between these two dynamics can make an adiabatic motion on the cavity-BO surface very fragile \cite{Galego2015}. This is indeed the case outside the cavity, where following the formation of the 1-exciton bright state energy localization is very fast \cite{Li2022b}, because this state is energetically embedded in a high density of dark states or, equivalently for our considerations, a large density of states in which one molecule is locally excited \footnote{The role of dark states in molecular polariton dynamics is often discussed \cite{Virgili2011,Galego2015,Eizner2019,Liu2020}. In room temperature molecular systems phase relations between molecules are likely quickly destroyed so a picture involving locally excited molecules is more realistic. As sinks for bright state relaxation the two bases are equivalent.}.

In the cavity, the energy gap between the polaritonic state makes the cavity-BO dynamics more robust. Indeed, as shown in Ref. \cite{Zeb2017} (see also Ref. \cite{Wu2016}), provided that the Rabi spitting $\Omega_R$ is larger than the single-molecule reorganization energy $E_R=\hbar\omega_v\lambda^2$ the ground electronic state of the molecules-cavity mode system is the lower polariton. Even then, this state is often not the lowest \textit{free energy state}, and at finite temperature its coupling to the much larger set of dark modes may disrupt adiabatic motion on the polariton surface. In another recent paper, Cederbaum \cite{Cederbaum2021} has analyzed in detail the potential energy surfaces of the lower and higher polaritons, as well as of dark states, in the vicinity of the uniformity curve (the subspace in which all molecules move uniformly along identical positions in their respective nuclear subspaces). These studies (see also the numerical studies of Refs. \cite{Luk2017,Hulkko2021} and the recent cavity-BO based study of Ref. \cite{Bonini2021}) provide important insights, but a fundamental understanding of the collective dynamics involving cavity polaritons is still lacking, making complete interpretation of experimental observations difficult.

On the experimental side, while a quantitative comparison of theories and experiments has not been so far reached, most observations of the effect of polariton excitation on subsequent molecular processes appear to be related to the polaritonic energy shift \cite{Eizner2019,Avramenko2020,Stranius2018,Yu2021,Esteso2021}), while manifestations of the predicted polaron decoupling effect are yet to be observed. The absence of clear experimental observation may be related to the aforementioned fragility.

As already mentioned, the dynamics under study are characterized by the co-existing of collective coupling of $N$ molecules with the radiation field, and internal dynamics that respond to the excitation state of individual molecules. In the present work we introduce and study a simplified analog of the HTC model in which the internal harmonic oscillators are replaced by 2-level systems. This implies a highly reduced dimensionality of the system Hilbert space compared to the HTC model. Further reduction, based on disregarding states that can be populated only at long times, makes it possible to study the short time behavior of this model by the direct diagonalization of the model Hamiltonian. Like the HTC model, this simplified version is characterized by the interplay of collective molecules-field interactions (determined by the parameters $g$ and $N$ in the Hamiltonian \eqref{eq:HHTC}) and local internal molecular dynamics (dominated by a coupling parameter $\lambda$ (see Sec. \ref{sec.2}) equivalent to the parameter $\lambda$ in the Hamiltonian \eqref{eq:HHTC}), and can be used to study the consequence of this interplay on different observables. We use it to examine the consequence of polariton (and the associated collective molecular bright state) excitation on subsequent internal molecular dynamical processes. The polaron decoupling picture, according to which the potential surface underlying every molecular nuclear motion in the polariton state reflects the bright state nuclear potential energy $((N-1)V_G(\mathbf{R})+V_X(\mathbf{R}))/N$ ($\mathbf{R}$ the nuclear configuration), would imply that in the polaritonic state the molecular internal dynamics will proceed under an effective coupling $\lambda/N$, where $N$ is the number of coupled molecules. Our simulations can examine this process directly.

Details of our model are provided in Sec. \ref{sec.2}, while Sec. \ref{sec.3} describes our numerical procedures. Results of our calculations are presented, compared and discussed in Sec. \ref{sec.4}. Section \ref{sec.5} concludes.

\section{Model}
\label{sec.2}
Consider the HTC model, Eq. \eqref{eq:HHTC}, without the intermolecular coupling
\begin{equation}
    \hat{H}_{HTC}=\hbar\omega_c\hat{a}^\dagger\hat{a}+\hbar\sum_{j=1}^N\left[\omega_{xg}\hat{\sigma}^+_j\hat{\sigma}^-_j+\frac{g}{2}(\hat{a}^\dagger\hat{\sigma}^-_j+\hat{\sigma}^+_j\hat{a})+\omega_v\hat{b}^\dagger_j\hat{b}_j+\lambda\hat{\sigma}^+_j\hat{\sigma}^-_j(\hat{b}^\dagger_j+\hat{b}_j)\right].
    \label{eq:HTC}
\end{equation}
In the present study, this Hamiltonian describes a system of $N$ 2-electronic-levels molecules with energy spacing $\hbar\omega_{xg}$ interacting with a cavity mode of frequency $\omega_c$, and through it with each other. In addition, each molecule undergoes internal harmonic motion about an equilibrium position that is shifted by $\lambda$ between the two electronic states. We simplify this model by replacing the molecular oscillators with two-level entities and the harmonic oscillations by Rabi oscillations between these levels. Denoting these inner states on molecule $j$ by $|a_j\rangle$ and $|b_j\rangle$ and defining $\hat{\tau}_j^+=|b_j\rangle\langle a_j|,\hat{\tau}_j^-=|a_j\rangle\langle b_j|$, the Hamiltonian is 
\begin{equation}
    \hat{H}=\hbar\omega_c\hat{a}^\dagger\hat{a}+\hbar\sum_{j=1}^N\left[\omega_{xg}\hat{\sigma}^+_j\hat{\sigma}^-_j+\frac{g}{2}(\hat{a}^\dagger\hat{\sigma}^-_j+\hat{\sigma}^+_j\hat{a})+\Delta\omega\hat{\tau}^+_j\hat{\tau}_j^-+\lambda\hat{\sigma}^+_j\hat{\sigma}^-_j(\hat{\tau}^+_j+\hat{\tau}_j^-)\right].
    \label{eq:ETC}
\end{equation}

Note that Eqs. \eqref{eq:HTC} and \eqref{eq:ETC} are of similar forms except that the parameters $\Delta\omega$ and $\lambda$ now characterize the two-level Hamiltonian that replaced the harmonic oscillator on each molecule. We will sometimes use the term “nuclear” for this motion, but the essential feature is that this is an internal molecular degree of freedom whose dynamics depends on the molecular electronic state: in the ground electronic state of molecule $j$ it is described by the Hamiltonian $\Delta\omega\hat{\tau}_j^+\hat{\tau}_j$, where $\Delta\omega=\epsilon_b-\epsilon_a$ denotes the energy spacing between the two internal levels, while in the excited electronic state it is $\Delta\omega\hat{\tau}_j^+\hat{\tau}_j^-+\lambda(\hat{\tau}_j^++\hat{\tau}_j^-)$. The parameter $\lambda$ is the analog of the coupling between the electronic and nuclear motion of the HTC model. Some more points should be noted:

(a) The model represented by the Hamiltonian \eqref{eq:ETC} is not aimed to describe a specific molecular process – only as a simplified tool for examining the implications of a collective electronic excitation (bright mode of a molecular ensemble or the corresponding polariton mode when this ensemble is in an optical cavity) on an internal molecular degree of freedom whose relatively slow dynamics depends on the molecular electronic state. These internal dynamics can be regarded as the analog of molecular nuclear motion, but could also represent a transition between an electronic state that is optically accessible from the ground electronic state and another state that is not. It should be kept in mind that in order to focus on the interplay between the collective molecules-cavity dynamics and internal molecular motions we have disregarded interactions between molecules that do not arise from their mutual coupling to the cavity [60].

(b) In the studies describe below, we limit ourselves to the single exciton subspace of the excited molecular system. In accordance, the cavity mode can be described as a two-level system, $|g_0\rangle$ and $|e_0\rangle$, so that $\hat{a}=|g_0\rangle\langle e_0|\equiv\hat{\sigma}_0^-$ and $\hat{a}^\dagger=|e_0\rangle\langle g_0|\equiv\hat{\sigma}_0^+$.

(c) As in the HTC model, the electronic coupling to the radiation field (cavity mode or the external field), represented by the parameter $g$, is assumed not to involve, namely not to change, the "internal" nuclear states. This is the analog of the Condon approximation in molecular spectroscopy. For specificity, we assume that all molecules start in the "internal" state $|a\rangle$ on the ground electronic state, so the initial internal state is $\prod_j|a_j\rangle$. In this case a sudden broad band electronic excitation moves it to the same state in the excited electronic state, thus initiating $|a_j\rangle\leftrightarrow|b_j\rangle$ dynamics (internal Rabi oscillations with frequency $\lambda$) between the two “internal” levels in the subspace in which molecule $j$ is excited. The question is, how will this internal molecular dynamics be expressed following the excitation of a polariton involving the bright electronic state of $N$ molecules.

(d) For $\Delta\omega=0$, the timescale associated with the internal $|a_j\rangle\leftrightarrow|b_j\rangle$ motion is characterized by the parameter $\lambda^{-1}$ (Rabi oscillations with frequency $\lambda$). In our implementation of the model \eqref{eq:ETC} we will usually take $\Delta\omega=0$ and $g\sqrt{N}>\lambda>g$, which is equivalent to the timescale ordering given by Eq. \eqref{eq:timediff}. For this case, the polaron decoupling picture would predict oscillations in the internal subspace with a frequency of order $N^{-1}\lambda$ because only one out of $N$ molecules is excited (and therefore experiences the coupling $\lambda$) in the excited collective state $|X_j\rangle$. This prediction is examined against the numerical calculations presented below. Additional insights can be obtained by allowing $\Delta\omega$ to be of order $\Omega_R=g\sqrt{N}$, as shown in Fig. \ref{fig:damp} and the related discussion.

(f) The present study focuses on the implication of collective polariton excitation on the subsequent unitary time evolution under the Hamiltonian \eqref{eq:ETC}. An important physical aspect – environmentally induced thermal relaxation, is not included and will be considered in a future study. In some of the studies (reported below) we consider the zero-temperature limit of such relaxation by coupling, within the internal subspace of each molecule $j$, either level $|a_j\rangle$ or level $|b_j\rangle$ to their own broad-band continua, representing a chemical reaction undergone by individual molecules following population of their $a$ or $b$ levels.

As is often done in studies of the HTC dynamics, we will use our simplified version \eqref{eq:ETC}
to describe the time evolution of a system in the single exciton subspace by assuming that the excitation is weak enough so that excitonic correlations can be disregarded. Together with the cavity mode, this implies dimensionality $N+1$ for the electronic subspace. With the model introduced for the internal molecular subspace (two “nuclear” states per molecule), the dimension of the full Hilbert space is $(N+1)2^N$, which, for large $N$, cannot be solved by direct diagonalization. We therefore invoke another basis truncation using the following consideration: let the initial state of the molecular system be $|G\rangle|V_0\rangle$ where 
\begin{equation}
    |G\rangle=\prod_j|g_j\rangle;\quad |V_0\rangle=\prod_j|a_j\rangle;\quad |G\rangle|V_0\rangle=\prod_j|g_ja_j\rangle
    \label{eq:GV0}
\end{equation}
(all molecules are in the ground electronic state $g$ and in the internal “nuclear” state $a$). Denote also the single exciton bright state by
\begin{equation}
    |B\rangle=N^{-\frac{1}{2}}\sum_{j=1}^N|X_j\rangle
\end{equation}
where $|X_j\rangle=|x_j\rangle\prod_{k\neq j}|g_k\rangle,j=1,...,N$ is a product of molecular electronic states with molecule $j$ excited and all others in their ground state. For the “internal” subspace, in addition to $|V_0\rangle$ we consider also the states 
\begin{equation}
    |V_j\rangle\equiv|b_j\rangle\prod_{k\neq j}|a_k\rangle.
    \label{eq:Vj}
\end{equation}
Out of the cavity, under our assumption that the optical interaction does not change the internal "nuclear" states, a one-photon excitation leads the vibronic bright state
\begin{equation}
    |B\rangle|V_0\rangle=N^{-\frac{1}{2}}\sum_{j=1}^N|X_ja_j\rangle\prod_{k\neq j}|a_k\rangle.
    \label{eq:brightvibro}
\end{equation}
We use the form on the right to emphasize that following this excitation, each term in the linear combination \eqref{eq:brightvibro} will oscillate between $|X_ja_j\rangle$ and $|X_jb_j\rangle$ with frequency $\lambda$. In this case (out of cavity) there is never more than one molecule in state $b$. The “nuclear” dynamics can be described in the basis of the $N+1$ states $V_j,j=0,...,N$ and the system dynamics can be described in the basis of only $2N$ states: $|X_j\rangle|V_0\rangle$ and $|X_j\rangle|V_j\rangle,j=1,...,N$. Inside the cavity $(g\neq 0)$ this is no longer true. Pathways such as
\begin{equation}
    |X_j\rangle|V_j\rangle\rightarrow|C\rangle|V_j\rangle\rightarrow|X_{j'}\rangle|V_j\rangle\rightarrow|X_{j'}\rangle|U_{jj'}\rangle\rightarrow|C\rangle|U_{jj'}\rangle\rightarrow...,
    \label{eq:XtoC}
\end{equation}
where $|C\rangle$ (henceforth also denoted $|X_0\rangle$) denotes the state in which the cavity mode is excited and all molecules are in their ground electronic state, and where $|U_{jj'}\rangle\equiv|b_j\rangle|b_{j'}\rangle\prod_{k\neq j,j'}|a_k\rangle$, become possible, eventually encompassing the full $2^N(N+1)$ dimensional Hilbert space. However, in contrast to the process $|X_j\rangle|V_0\rangle\leftrightarrow|C\rangle|V_0\rangle$ which represents the collective Rabi oscillation between the initial vibronic bright state Eq. \eqref{eq:brightvibro} and the cavity and therefore takes place on the timescale $(g\sqrt{N})^{-1}$, the $X\leftrightarrow C$ processes in Eq. \eqref{eq:XtoC} are individual molecular processes whose characteristic timescale is of order $g^{-1}$. Limiting our analysis to time that can be long relative to the collective Rabi period $(g\sqrt{N})^{-1}$ but short relative to $g^{-1}$ makes it possible to use, for large enough $N$, a truncated basis that comprises the states $|X_j\rangle|V_k\rangle, j,k=0,...,N$ (note again that $|C\rangle=|X_0\rangle$), making it a basis of order $(N+1)^2$. As a shorthand notation we also use below $|j,k\rangle\equiv|X_j\rangle|V_k\rangle$. In addition to the states involved in the out of cavity dynamics, this basis includes states with $j$ and $k$ different from zero and also different from each other. Such states cannot be populated out of the cavity and become populated in the cavity on the timescale $\sim g^{-1}$. The performance of this approximation (and the rate of its deterioration over time) can be therefore monitored by following the population of such states.

The interplay between collective and local dynamics can be explicitly described within this model. Both in the HTC model and in our simplified version, every vibronic state of the form $|B\rangle|V\rangle$ (a product of the molecular electronic bright state a molecular internal "nuclear", inner state) is coupled to the cavity state $|C\rangle|V\rangle$ to form two vibronic polaritons. Our model assumption is that the ground vibronic state of the molecular system is coupled only to the state $|B\rangle\prod_j|a_j\rangle$ and therefore to the vibronic polaritons $2^{-1/2}(|C\rangle\pm|B\rangle)\prod_j|a_j\rangle$. However, the “vibronic coupling” term $\lambda\sum_{j=1}^N\hat{\sigma}_j^+\hat{\sigma}_j^-(\hat{\tau}_j^++\hat{\tau}_j^-)$ couples the state $|B\rangle\prod_j|a_j\rangle$ to states which are not of the form of a product $|B\rangle|V\rangle$
\begin{equation}
    \lambda\sum_{j=1}^N\hat{\sigma}_j^+\hat{\sigma}_j^-|B\rangle\prod_k(\hat{\tau}_j^++\hat{\tau}_j^-)|a_k\rangle=\frac{\lambda}{\sqrt{N}}\sum_{j=1}^N|X_j\rangle|b_j\rangle\prod_{k\neq j}|a_k\rangle.
    \label{eq:BV}
\end{equation}
The states on the RHS of Eq. \eqref{eq:BV} cannot couple collectively to the same vibronic-cavity state, and for large $N$ they contribute mostly to the density of dark states. If $\Delta\omega=0$ in the Hamiltonian \eqref{eq:ETC}, this implies that the vibronic coupling $\lambda\sum_{j=1}^N\hat{\sigma}_j^+\hat{\sigma}_j^-(\hat{\tau}_j^++\hat{\tau}_j^-)$ couples the polaritonic states $2^{-1/2}(|C\rangle\pm|B\rangle)\prod_j|a_j\rangle$ to states whose energy is higher (for the lower polariton) or lower (for the upper polariton) by nearly a half collective Rabi splitting $\Omega_R/2$. This results in an apparent conflict: energy conservation implies that the evolution of an initially excited polariton state $2^{-1/2}(|C\rangle\pm|B\rangle)\prod_j|a_j\rangle$ will mostly remain in the subspace of states of the form $2^{-1/2}(|C\rangle\pm|B\rangle)|V\rangle$, forcing the nuclei to move collectively, while the vibronic coupling that induces this evolution does not directly couple such state. Thermal interactions at finite temperature will bridge the energy gap and, because of the large number of dark states, promote the importance of entropic considerations. Here, however, we focus on the way the system resolves this conflict in the absence of thermal interactions.  

\section{Numerical Procedures}
\label{sec.3}
In the truncated basis $|j,k\rangle\equiv|X_j\rangle|V_k\rangle$ the Hamiltonian \eqref{eq:ETC} is represented by an $(N+1)^2\times(N+1)^2$ matrix that is the tensor product of $N+1$-dimensional electronic subspace matrix (1-exciton Tavis Cummings Hamiltonian) spanned by the states $|X_j\rangle,j=0,...,N$ and the $N+1$-dimensional matrix representing the molecular inner subspace spanned by the states $|V_k\rangle,k=0,...,N$, 
\begin{align}
    \mathbf{H}&=\left(\begin{matrix}
    \mathbf{M}^{(00)} &\mathbf{M}^{(01)} &... &\mathbf{M}^{(0N)}\\
    \mathbf{M}^{(10)} &\mathbf{M}^{(11)} &... &\mathbf{M}^{(1N)}\\
    &...\\
    \mathbf{M}^{(N0)} &\mathbf{M}^{(N1)} &... &\mathbf{M}^{(NN)}\\
\end{matrix}\right).
 \label{eq:matrixETC}
\end{align}
Each of the matrices $\mathbf{M}$ is of order $N+1$. On the diagonal, we have, for $j=0,...,N$,
\begin{align}
    \hbar^{-1}\mathbf{M}^{(jj)}&=\left(\begin{matrix}
    \omega_c+\Delta \omega(1-\delta_{j0}) &\frac{g}{2} &\frac{g}{2} &... &\frac{g}{2}\\
    \frac{g}{2} &\omega_{xg}+\Delta \omega(1-\delta_{j0}) &0 &... &0\\
    \frac{g}{2} &0 &\omega_{xg}+\Delta \omega(1-\delta_{j0}) &... &0\\
    &...\\
    \frac{g}{2} &0 &... &0 &\omega_{xg}+\Delta \omega(1-\delta_{j0})
\end{matrix}\right).
\end{align}
All elements of the matrices $\mathbf{M}^{jj'}$ with $j\neq j'$ are 0 if both $j,j'$ are non zero, while the matrices $\mathbf{M}^{(0j)}$ and $\mathbf{M}^{(j0)}, j=1,...,N,$ have only one non-zero element  $\mathbf{M}^{(0j)}_{kl}=\mathbf{M}^{(j0)}_{kl}=\lambda\delta_{k,j+1}\delta_{l,j+1}$. The submatrices $\mathbf{M}^{(jj)}$ represent pure electronic TC dynamics while $\mathbf{M}^{(0j)}$ and $\mathbf{M}^{(j0)}$ affect inner state coupling that does not change the TC part.

The effect of collective excitation on subsequent system evolution can be studied using this model or its variants in several modes of calculations.

\textbf{Mode A}.
Focusing on the case were the cavity mode frequency exactly matches the molecular transition frequency, $\omega_c=\omega_{xg}$, we assume that the system is initially prepared, when out of cavity, in the bright state $|\Psi(0)\rangle=|B\rangle|V_0\rangle$, or, inside the cavity, in the polariton states derived from this bright state 
\begin{align}
   |\Psi(0)\rangle&=\frac{1}{\sqrt{2}}\left(|X_0\rangle\pm\frac{1}{\sqrt{N}}\sum_{j=1}^N|X_j\rangle\right)|V_0\rangle.
   \label{eq:initial}
\end{align}
By our assumptions, in these states all molecules occupy their internal state $a$. Here and below we refer to the $+/-$ states in Eq. \eqref{eq:initial} as the upper/lower $a$-polariton states. Following this preparation, the system propagated according to $\Psi(t)=\exp(-i\hat{H}t/\hbar)\Psi(0)$ is evaluated by diagonalizing the truncated Hamiltonian. This yields
\begin{align}
    |\Psi(t)\rangle=\sum_{j=0}^N\sum_{k=0}^Nc_{jk}(t)|j,k\rangle.
    \label{eq:wavf}
\end{align}
The following observables are then evaluated:
\begin{align}
    r_0(t)&=|c_{00}(t)|^2
    \label{eq:r0t}
\end{align}
and
\begin{align}
    r_1(t)&=\sum_{k=1}^N|c_{0k}(t)|^2.
    \label{eq:r1t}
\end{align}
\label{eq:rt}
are the probabilities that at time $t$ the cavity mode is excited and the molecular subsystem is in the initial $|V_0\rangle$ or any other $|V_k\rangle,k=1,...,N$ inner states, respectively. Similarly,
\begin{equation}
    d_0(t)=\sum_{j=1}^N|c_{j0}(t)|^2
    \label{eq:d0t}
\end{equation}
is the probability that the molecular system excited at time $t$ is in its initial inner state $|V_0\rangle$ in which all molecules are in their internal state $a$;
\begin{align}
    d_1(t)&=\sum_{j=1}^N|c_{jj}(t)|^2,
    \label{eq:d1t}
\end{align}
is the probability the molecular system is in states where the molecule that is excited has changed its inner state and
\begin{align}
    d_2(t)&=\sum_{j=1}^N\sum_{k(\neq j)=1}^N|c_{jk}(t)|^2,
    \label{eq:d2t}
\end{align}
is the probability that in the molecular system a molecule that is not excited has moved to another inner state. Note that out of cavity the truncation is not an approximation for the chosen initial state. In the cavity, $d_2(t)$ increases slowly in time (shown in the next section), and a considerable deviation from zero can be used as an indicator of possible failure of the truncated basis dynamics.

\textbf{Mode B}.
The quantities defined in Eqs. (\ref{eq:r0t}-\ref{eq:d2t}) are numerical indicators of system dynamics that may be affected by the collective nature of the initial state. An actual observable may be related to a process that occurs following the molecular excitation and/or the internal $a\leftrightarrow b$ dynamics. To build such processes into the present model we impose decay pathways on different molecular levels. For example, dissociation to form a product $P$ out of states $b$, following the transition $a\rightarrow b$ can be implemented in our model by augmenting only the $j+1$-th diagonal element of the matrices $\mathbf{M}^{(jj)}$ with $j\geq 1$ by an imaginary term $-(i/2)\eta_b$. The instantaneous rate of product formation is then
\begin{equation}
    \frac{dP}{dt}=\eta_b\sum_{j=1}^N|c_{jj}(t)|^2,
    \label{eq:dPt}
\end{equation}
while the average product formation time (inverse production rate) may be estimated from loss of normalization due to escapes into the product space
\begin{equation}
    \tau_P=\int_0^\infty dt |\langle\Psi(t)|\Psi(t)\rangle|,
    \label{eq:prate}
\end{equation}
where $\Psi(t)$ is given by Eq. \eqref{eq:wavf}.
The absolute magnitude of this time should be regarded with caution: it is meaningful only provided it is smaller than the time at which the calculation based on our truncated basis becomes invalid. Still, as we will see below, the dependence of $\tau_P$ on system parameters, including the out of cavity $(g=0)$ behavior, can help analyze the interplay between collective excitation and subsequent product formation due to a process undergone by individual molecules.

Note that a simple model in which collective and local relaxation dynamics come into play can be constructed starting from the TC Hamiltonian $\hat{H}_{TC}=\hbar\omega_c\hat{a}^\dagger\hat{a}+\hbar\sum_{j=1}^N[\omega_{xg}\hat{\sigma}_j^+\hat{\sigma}_j^-+g(\hat{a}^\dagger\hat{\sigma}_j^-+\hat{a}\hat{\sigma}_j^+)/2]$, the vibrationless analogue of Eq. \eqref{eq:HTC}, and replacing $\omega_{xg}$ by $\omega_{xg}-i\eta/2$. Here, $\eta$ represents the decay rate of an individual excited molecule. It is easily realized that the corresponding lifetime of the bright state of this molecular ensemble (excited outside the cavity), is not affected by its collective nature and remains $\eta^{-1}$, same as that of an individually excited molecule. Similarly in the cavity, the lifetime of each polariton reflects its molecular component and is $(\eta/2)^{-1/2}$ in the symmetric case $\omega_c=\omega_{xg}$. In contrast, we will see below that, the rate of product formation following the $a\rightarrow b$ transition, calculated from Eq. \eqref{eq:dPt}, does reflect the collective nature of the initially excited state.

\textbf{Mode C}.
A simple variant of the model \eqref{eq:ETC} makes it possible to study the response of our system to CW excitation. This variant is obtained from imposing additional driving and damping terms to the model Hamiltonian \eqref{eq:ETC} (in which the cavity mode is represented as a 2-state entity) as follows (see Sec. II in the SI for details). First, a driving state of frequency $\omega$ (representing the system zero-energy ground state dressed by a far-field photon of frequency $\omega$) that is coupled to the cavity mode and/or to the molecular bright state, $|B\rangle|V_0\rangle$, with a coupling amplitude $W$ (assumed small enough to justify linear response consideration). Second, each state in the excited state manifold of the uncoupled system (represented by the Hamiltonian of Eq. \eqref{eq:ETC} without the $g,\lambda$ terms) is assumed to be coupled to its own broad band continuum, implying that the evolution equations for the coefficients $c_{jk}(t)$ of Eq. \eqref{eq:wavf} are modified by adding damping terms, $-\eta_{jk}c_{jk}/2$. In terms of the Green's "function" associated with the Hamiltonian of Eq. \eqref{eq:matrixETC} 
\begin{equation}
    \mathbf{G}=\frac{1}{\omega\mathbf{I}-\mathbf{H}+i\pmb{\eta}/2},
    \label{eq:G}
\end{equation}
where $\mathbf{I}$ is the unit matrix and $\pmb{\eta}$ is the diagonal matrix of damping coefficients, this leads (see Sec. II in the SI) to the following expressions for the absorption lineshape and yields
\begin{align}
    \label{eq:lineshape}
    L(\omega)&=-\text{Im}G_{00,00}(\omega),\\
    Y_{jk}(\omega)&=-\frac{\eta_{jk}|G_{00,jk}|^2}{\text{Im} G_{00,00}}.
\end{align}
Note that while formally we can impose a damping channel on any state of the system, in our practical implementations we impose three damping parameters. A damping parameter $\eta_0$ is imposed on all states $|0,k\rangle$ in which the cavity mode is excited. It represents all possible decay channels of this mode: reflection, transmission and dissipation in the cavity boundaries. Second, on the molecular side, all states $|j,j\rangle, (j=1,...,N)$ in which a molecule is electronically excited and the same molecule is in inner state $b$, are assigned a “reactive” decay (product formation rate) $\eta_b$, while all other states ($|j,k\rangle$ with $j=1,...,N,k=0,...,N$ but $k\neq j$, where molecule $j$ is electronically excited but not in internal state $b)$ are assigned a damping parameter $\eta_a$ that represents molecular relaxation processes that do not lead to an observed product. The "product formation yield" is then  
\begin{equation}
    Y_g(\omega)=-\eta_b\frac{\sum_{j=1}^N|G_{00,jj}|^2}{\text{Im}G_{00,00}}.
    \label{eq:yield}
\end{equation}

It is important to note that while the results \eqref{eq:lineshape}-\eqref{eq:yield} are exact for our model, their applicability to our physical problem is severely limited because the use of a truncated basis implies a short time approximation. The calculation of CW lineshape and yields  may still hold provided that the imposed damping  rates are large enough to sufficiently suppress population on that part of the system state space that we are not accounting for. Our effort to provide this validation is described below.

\textbf{Weighted density of states}. In addition to the dynamical calculations described above, more insights into the system behavior can be gained by examining the states that are accessible from the initial state. For the presently studied closed system, for the given initial state $\Psi(t=0)$ of Eq. \eqref{eq:initial}, the subsequent dynamics is obtained in terms of the eigenstates and corresponding energies, $\{\phi_n,\hbar\omega_n\}$, $\Psi(t)=\sum_n\langle\phi_n|\Psi(t=0)\rangle\phi_ne^{-i\omega_nt}$. The weight of some state $\Phi$ in the function $\Psi(t)$ is determined from the amplitude $\langle\Phi|\Psi(t)\rangle=\sum_n\langle\phi_n|\Psi(t=0)\rangle\langle\Phi|\phi_n\rangle e^{-i\omega_nt}$ and the weighted density (power spectrum of this weight) can be defined by \footnote{In our finite system these spectra are discrete, and we use instead coarse‐grained or smoothed forms, e.g., 
\begin{equation}
    D_\Phi(\omega)=\frac{1}{\delta\omega}\int_{\omega-\delta\omega/2}^{\omega+\delta\omega/2}\bar{D}_\Phi(\omega)d\omega
    \label{eq:DOS}
\end{equation}}
\begin{equation}
    \bar{D}_\Phi(\omega)=\sum_n|\langle\Phi|\phi_n\rangle|^2|\langle\phi_n|\Psi(t=0)\rangle|^2\delta(\omega-\omega_n).
\end{equation}
The choice of $\Phi$ is made so as to focus on any desired information, and an average over functions $\Phi$ in a desired range is also possible. In the calculations reported below we show the smoothed forms of the total density of states, $D_{tot}(\omega)=\sum_n\delta(\omega-\omega_n)$, then focus on the following weighted densities
\begin{subequations}\label{eq:wDOS}
\begin{align}
    D_1(\omega)&=\sum_n|\langle\phi_n|\Psi(t=0)|^2\delta(\omega-\omega_n)=D_2(\omega)+D_3(\omega),\\
    D_2(\omega)&=\sum_{j=0}^N\sum_n|\langle j0|\phi_n\rangle|^2|\langle\phi_n|\Psi(t=0)\rangle|^2\delta(\omega-\omega_n),\\
    D_3(\omega)&=\sum_{j=0}^N\sum_{k=1}^N\sum_n|\langle jk|\phi_n\rangle|^2|\langle\phi_n|\Psi(t=0)\rangle|^2\delta(\omega-\omega_n).
\end{align}
\label{eq:wDOS}
\end{subequations}
Accordingly, $D_1(\omega)$ is the density of states weighted by their projection on the initial state \eqref{eq:initial} while $D_2(\omega)$ and $D_3(\omega)$ are additionally weighted by their projection on the subspace of states in which no internal $a\rightarrow b$ transition has taken place and the complementary subspace for which such transition has occurred. Comparing these relative weights for molecular ensembles in and out of the cavity is another way to assess the local $a\rightarrow b$ dynamics following collective excitations in these environments.

\section{Results and discussion}
\label{sec.4}
The focus of the present study are the manifestation of the molecular collective response rather than spectroscopy. Therefore, unless otherwise stated, the results presented below are obtained for the case where the cavity mode frequency is the same as the single molecule electronic transition frequency $\omega_c=\omega_{xg}$ and the dependence on the number $N$ of involved molecules is shown for a given fixed Rabi splitting $\Omega_R=g\sqrt{N}$ so that $g=\Omega_R/\sqrt{N}$. More results showing the dependence on $\Omega_R$ at fixed radiative coupling $g$, are given in Sec. III of the SI. In the calculations reported below, we set $\hbar=1$ and present the energy in units of $\lambda$ (or $\lambda_0$, when $\lambda$ is distributed randomly about $\lambda_0$). Accordingly, the time unit is $1/\lambda$.

\begin{figure}[t]
\includegraphics[width=0.8\textwidth]{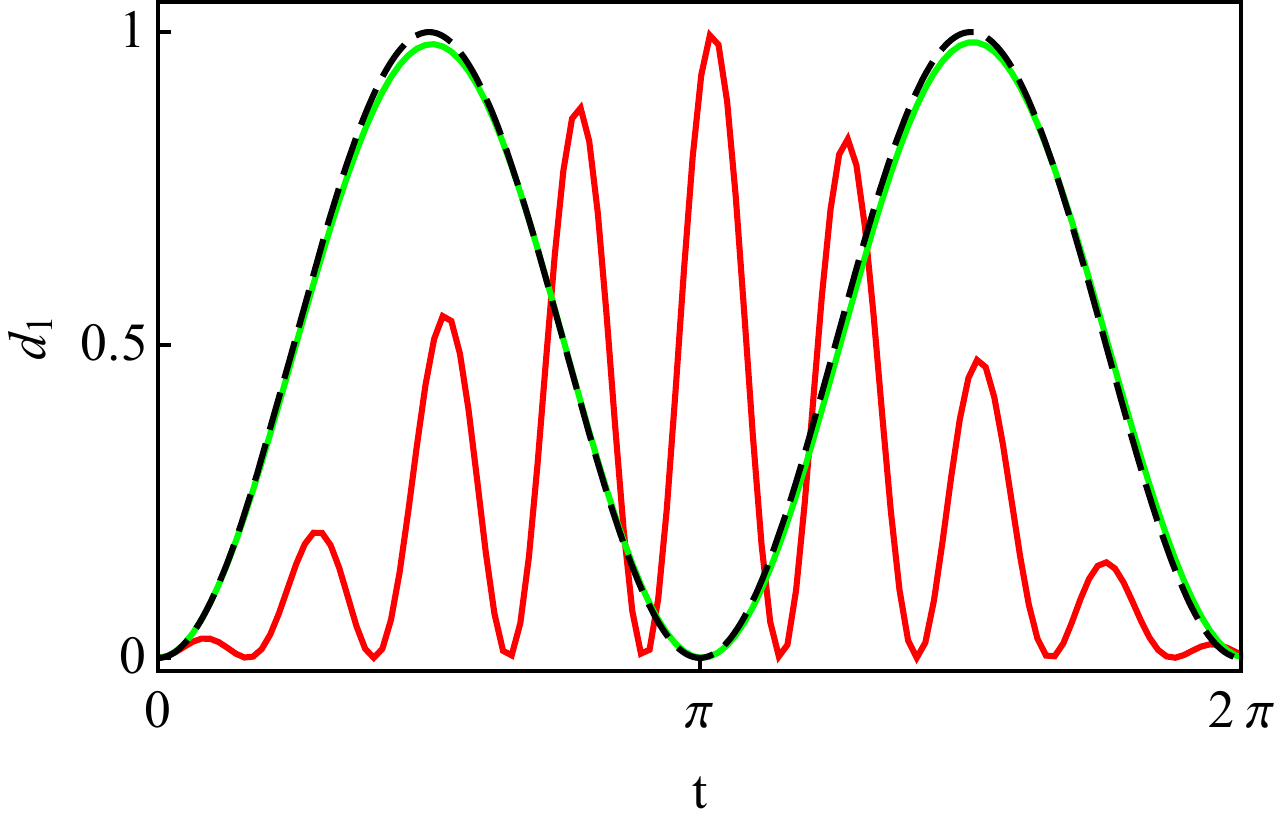}
\caption{The time evolution of population in internal state(s) $b$, given by $d_1(t)$ (Eq. \eqref{eq:d1t}), when the initial state is given by Eq. \eqref{eq:inimj1} (only one molecule is excited). Red and green lines correspond to systems of size $N=1$ and $N=120$, respectively; the black dashed line is the analytical result when out of cavity, given by $d_1(t)=\sin^2(\lambda t)$. The near overlap of the green and black dashed lines shows that for large $N$ a single excited molecule is not sensitive to the cavity environment. For the case $N=120$, the small magnitude of $d_2(t)$ (Fig. S2 in the SI) shows that on the timescale considered, other molecules are practically decoupled from the observed dynamics. Parameters are $\Delta \omega=0$, $\lambda=1$ and $g$ adjusted to keep  $\Omega_R=\sqrt{60}$.}
\label{fig:onlym1}
\end{figure}

For reference, consider first the case in which at $t=0$ only one molecule, say molecule 1, is excited with all molecules in inner state $a$,
\begin{equation}
    \Psi(t=0)=|X_1\rangle\prod_{k=1}^N|a_k\rangle.
    \label{eq:inimj1}
\end{equation}
Outside the cavity each molecule evolves individually, so
\begin{equation}
   |\Psi(t)\rangle=(\cos(\lambda t)|X_1a_1\rangle+\sin(\lambda t)|X_1b_1\rangle)\prod_{k\neq 1}|a_k\rangle,
\end{equation}
namely, $|c_{10}(t)|^2=\cos^2(\lambda t)$ and $|c_{11}(t)|^2=\sin^2(\lambda t)$, do not depend on $N$. Figure \ref{fig:onlym1} shows the corresponding time evolution for the same initial state in the cavity for the cases $N=1$ and $N=120$ (note that in the latter case only a single molecule is assumed to be excited at $t=0$) with $\Omega_R=\sqrt{60}$ in both cases. The effect of coupling to the cavity mode is clearly seen for the case $N=1$, however for large $N$ and during the time range $\Omega_R^{-1}<t<g^{-1}$ the internal dynamics of the initially excited molecule is similar to that seen outside the cavity. This behavior reflects the analysis provided in Sec. I in the SI: the coupling of a single excited molecule to other molecules in the cavity, due to their mutual coupling to the cavity mode, is of order $g^2\sim N^{-1}$ and effectively vanishes in the $N\rightarrow\infty$, $\Omega_R=$constant limit.

 When the single exciton bright state, Eq. \eqref{eq:brightvibro}, is initially excited outside the cavity the subsequent independent evolution of each molecule leads to
\begin{equation}
    \Psi(t)=N^{-1/2}\sum_{j=1}^N(\cos(\lambda t)|X_ja_j\rangle+\sin(\lambda t)|X_jb_j\rangle)\prod_{k\neq j}|a_k\rangle.
\end{equation}
It follows, using the definitions (\ref{eq:d0t}-\ref{eq:d1t}), that $d_0(t)=\cos^2(\lambda t)$ and $d_1(t)=\sin^2(\lambda t)$, showing that the internal dynamics of the bright state is the same as that of a single excited molecule. In contrast, when the corresponding polaritons, Eq. \eqref{eq:initial}, are populated at $t=0$ inside the cavity, the subsequent dynamics strongly depend on the cavity environment and on $N$, as demonstrated below. 

\begin{figure}[t]
\subfloat[][]{
\begin{minipage}[t]{0.5\textwidth}
\flushleft
\includegraphics[width=0.8\textwidth]{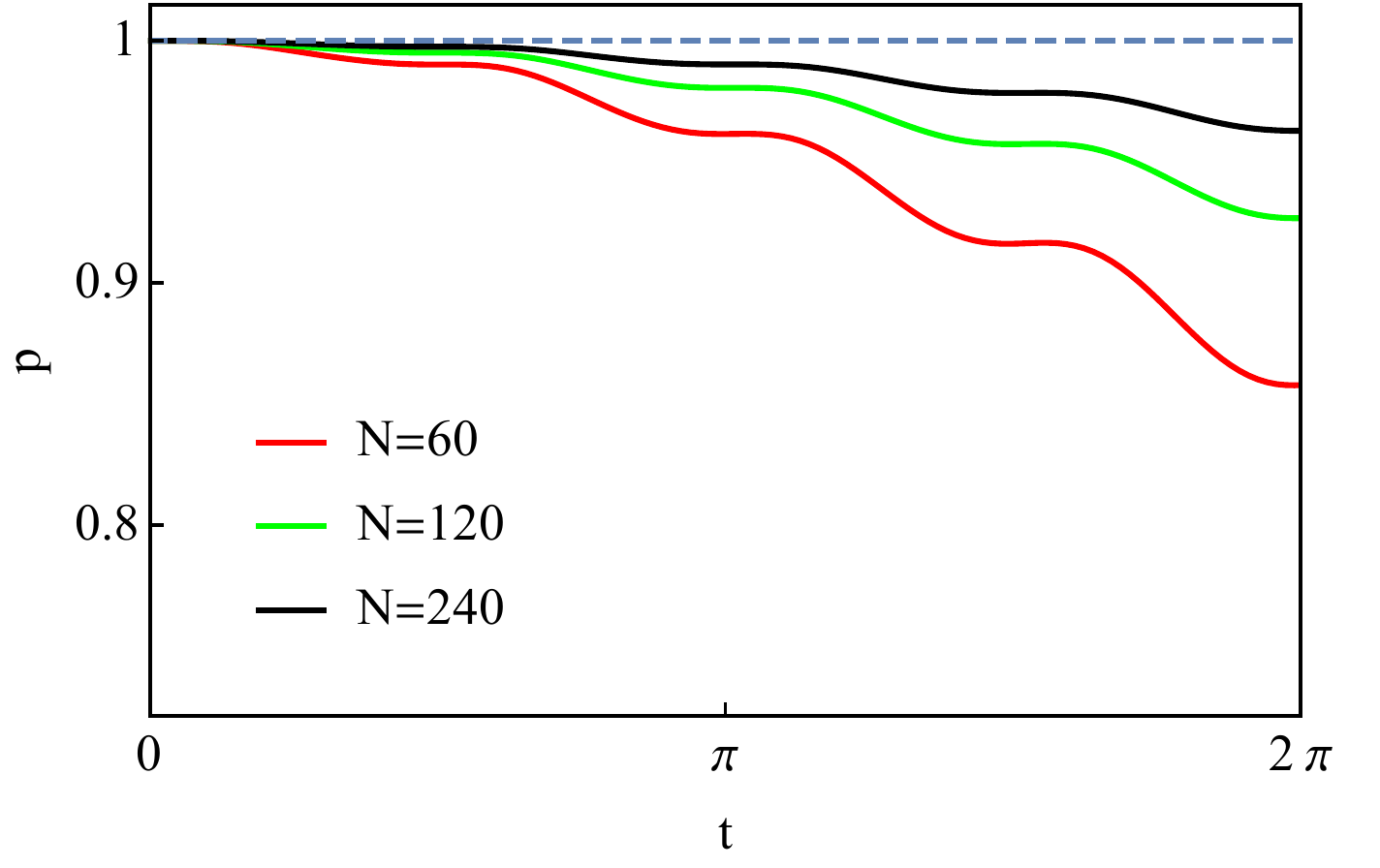}
\end{minipage}
}
\\
\subfloat[][]{
\begin{minipage}[t]{0.5\textwidth}
\flushleft
\includegraphics[width=0.8\textwidth]{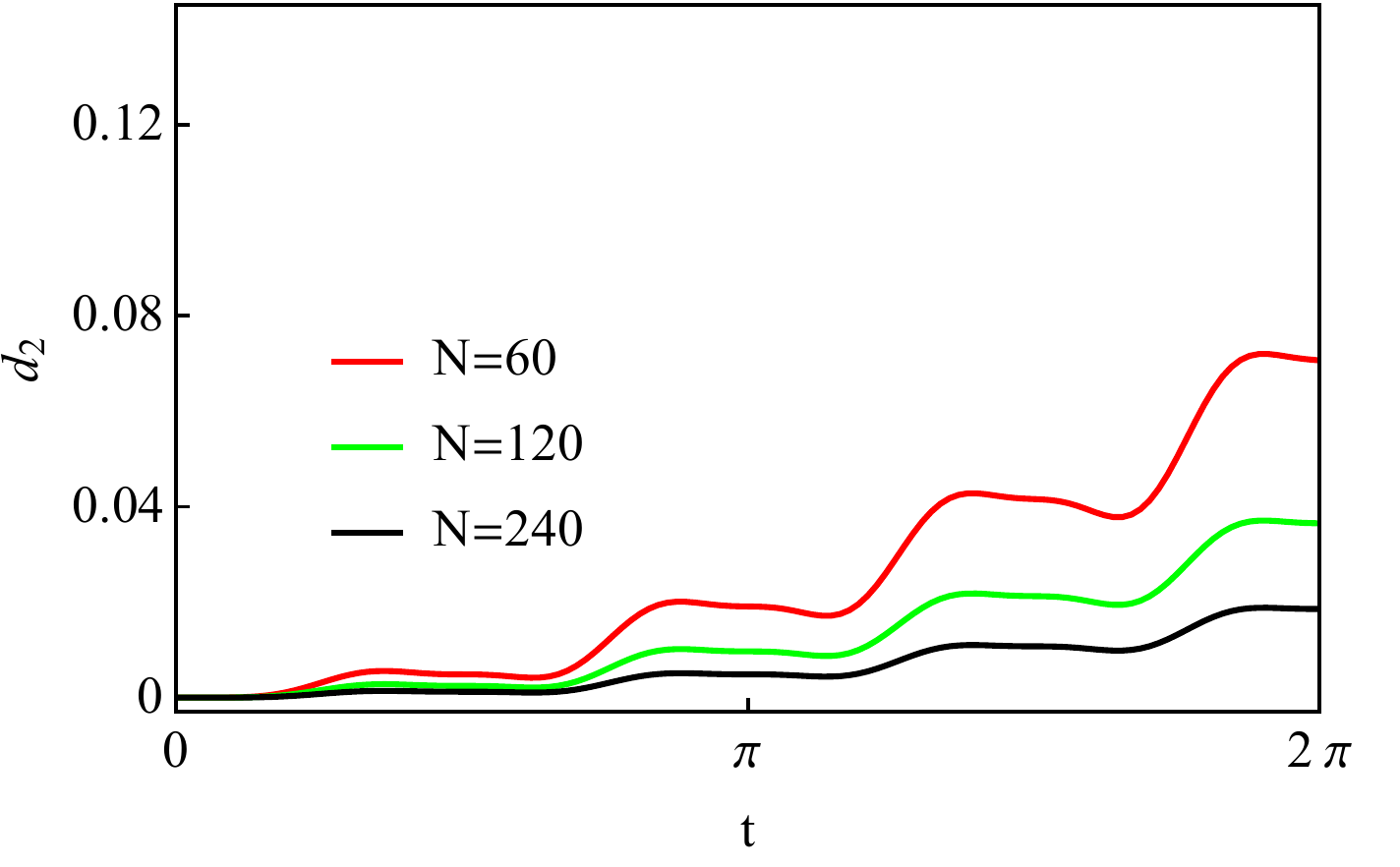}
\end{minipage}
}
\caption{Time evolution of (a) $p(t)$ and (b) $d_2(t)$, when the system starts from the lower $a$-polariton state, Eq. \eqref{eq:initial}. Red, green and black lines correspond to systems of size $N=60,120$ and $240$, respectively, adjusted so as to keep the collective Rabi frequency $\Omega_R=g\sqrt{N}$ constant $(\sqrt{60})$. The horizontal dashed line in panel (a) represents the $\lambda=0$ case, which is equivalent to the TC model. Parameters are $\omega_c=\omega_{xg}=0,\lambda=1$, $\Delta\omega=0$.}
\label{fig:shorttime}
\end{figure}

Before presenting these results we first examine in Fig. \ref{fig:shorttime} the validity of calculations based on our truncated basis during the timescale of our observation. This is done by observing the variables $d_2(t)$ and $p(t)=r_0(t)+d_0(t)+d_1(t)=1-r_1(t)-d_2(t)$ (cf. Eqs. (\ref{eq:r0t}-\ref{eq:d2t})). As discussed in Section \ref{sec.2}, starting from the $a$-polariton state (all molecules are in internal state $a$) and evolving under the Hamiltonian \eqref{eq:ETC}, populating a state $|U_{jj'}\rangle\equiv|b_j\rangle|b_{j'}\rangle\prod_{k\neq j,j'}|a_k\rangle$ in which two molecules are in internal state $b$ can take place only from states that contribute to $d_2$, while the latter can be populated only from states that contribute to $r_1$. Figure \ref{fig:shorttime} shows that for large $N$, $d_2(t)$ remains small on the timescale of our calculation \footnote{To understand the trend seen in Fig. \ref{fig:shorttime}, where $d_2$ is seen to grow more slowly with large $N$, we may look at one possible state of the $d_2$ set where the state $b$ is occupied on molecule $j$ while another molecule $k$ is excited, and follows the dynamics that will lead to its population. Starting from the bright state $N^{-1/2}\sum_j|X_j\rangle |V_0\rangle$ such a state will be obtained (the last state following step 3) by the sequence of transitions
\begin{align}
    N^{-1/2}\sum_j|X_j\rangle |V_0\rangle \stackrel{1}{\rightarrow} N^{-1/2}\sum_j|X_j\rangle|V_j\rangle\stackrel{2}{\rightarrow} |C\rangle|G\rangle|V_j\rangle\stackrel{3}{\rightarrow}N^{-1/2}\sum_k|X_k\rangle|V_k\rangle,
\end{align}
where $V_0$ and $V_j$ are defined by Eqs. \eqref{eq:GV0} and \eqref{eq:Vj}. The coupling for this sequence is the product of coupling for step 1 (order $N^{-1}$ because of the normalization), step 2 (order $N^{-1/2}$) and step 3 (order $N\cdot N^{-1/2}$, this is a collective step). The overall coupling is then of order $N^{-1}$ and the rate will be of order $N^{-2}$ multiplied by the number $N$ of possible final states $V_j$, which yields a rate of order $N^{-1}$. This dependence on $N$ is the reason why $d_2$ grows more slowly for large $N$ and why our truncation approximation will be valid for longer time for large $N$.}, thus providing support for the validity of results obtained using the truncated basis on the timescale considered.

\begin{figure}[t]
\subfloat[][]{
\begin{minipage}[t]{0.5\textwidth}
\flushleft
\includegraphics[width=0.8\textwidth]{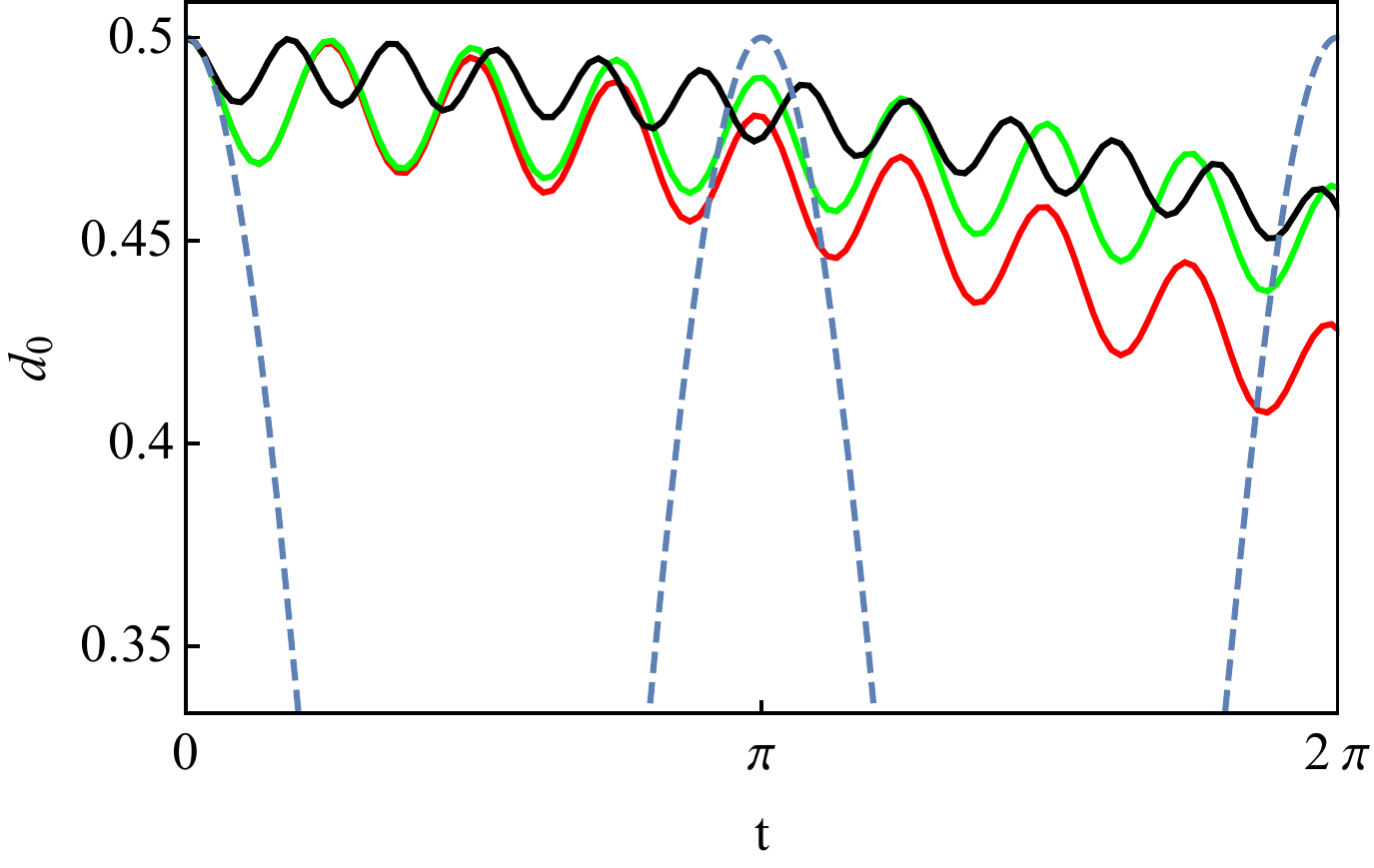}
\end{minipage}
}
\\
\subfloat[][]{
\begin{minipage}[t]{0.5\textwidth}
\flushleft
\includegraphics[width=0.8\textwidth]{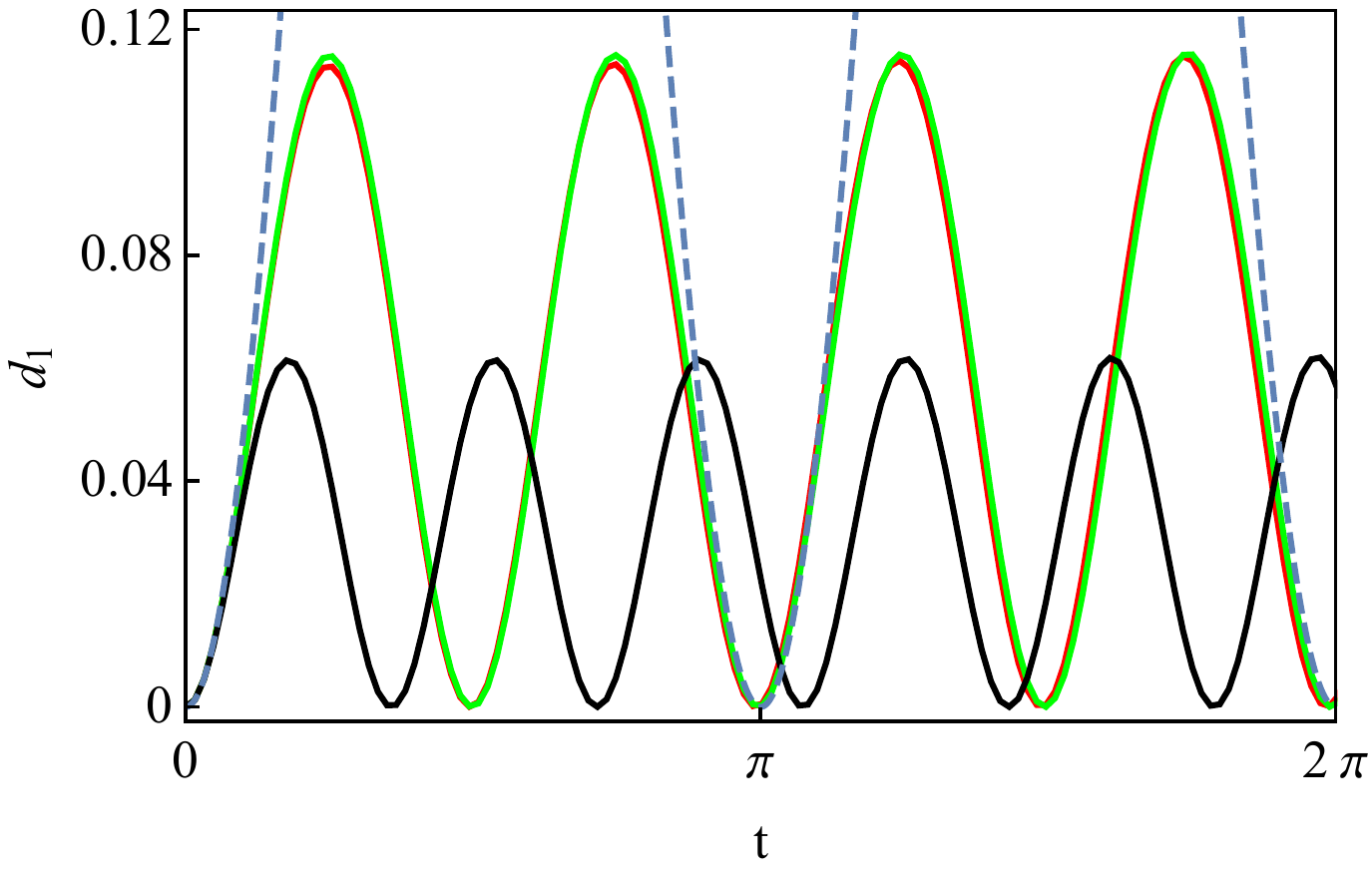}
\end{minipage}
}
\\
\subfloat[][]{
\begin{minipage}[t]{0.5\textwidth}
\flushleft
\includegraphics[width=0.8\textwidth]{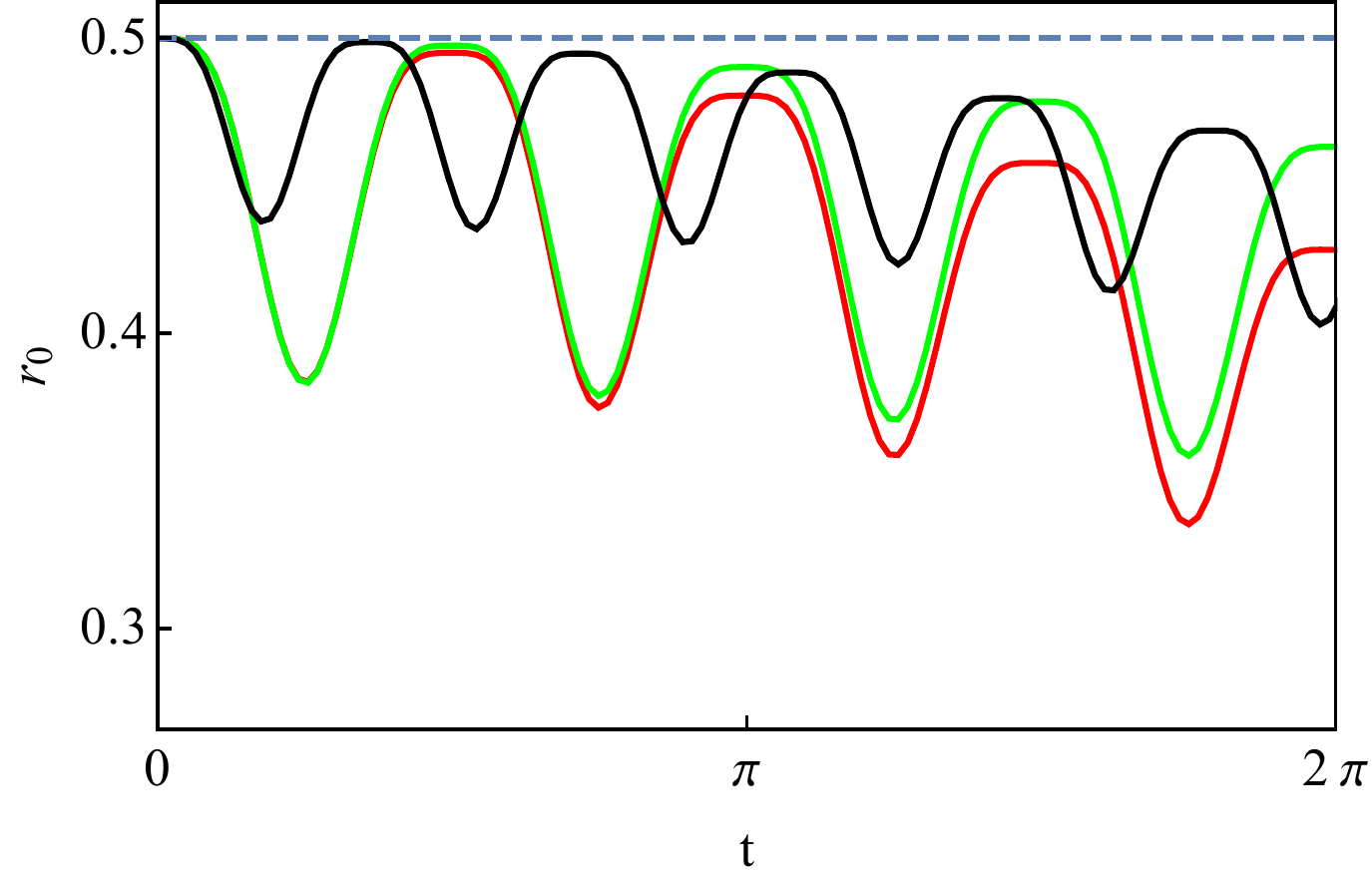}
\end{minipage}
}
\caption{Time evolution following excitation of the lower $a$-polariton. Plotted against time are, (a) the probability to stay in the all-$a$ subspace, $d_0(t)$, (b) the cumulative probability, $d_1(t)$, that excited molecules make the transition to the $b$ state, and (c) the probability $r_0$ that the cavity mode is populated while all molecules are in the ground electronic state and in their internal state $a$. Red, green and black lines correspond to systems with $N=60, \Omega_R=\sqrt{60}$; $N=120, \Omega_R=\sqrt{60}$ and $N=120, \Omega_R=\sqrt{120}$, respectively. The internal coupling parameter is $\lambda=1$. For comparison, also shown (blue dashed lines) are results $g=0$, starting from the same initial state Eq. \eqref{eq:initial}, which essentially represents dynamics out of cavity. Parameters are $\omega_c=\omega_{xg}=0$, $\Delta\omega=0$. }
\label{fig:collective}
\end{figure}

\begin{figure}[t]
\subfloat[][]{
\begin{minipage}[t]{0.5\textwidth}
\flushleft
\includegraphics[width=0.8\textwidth]{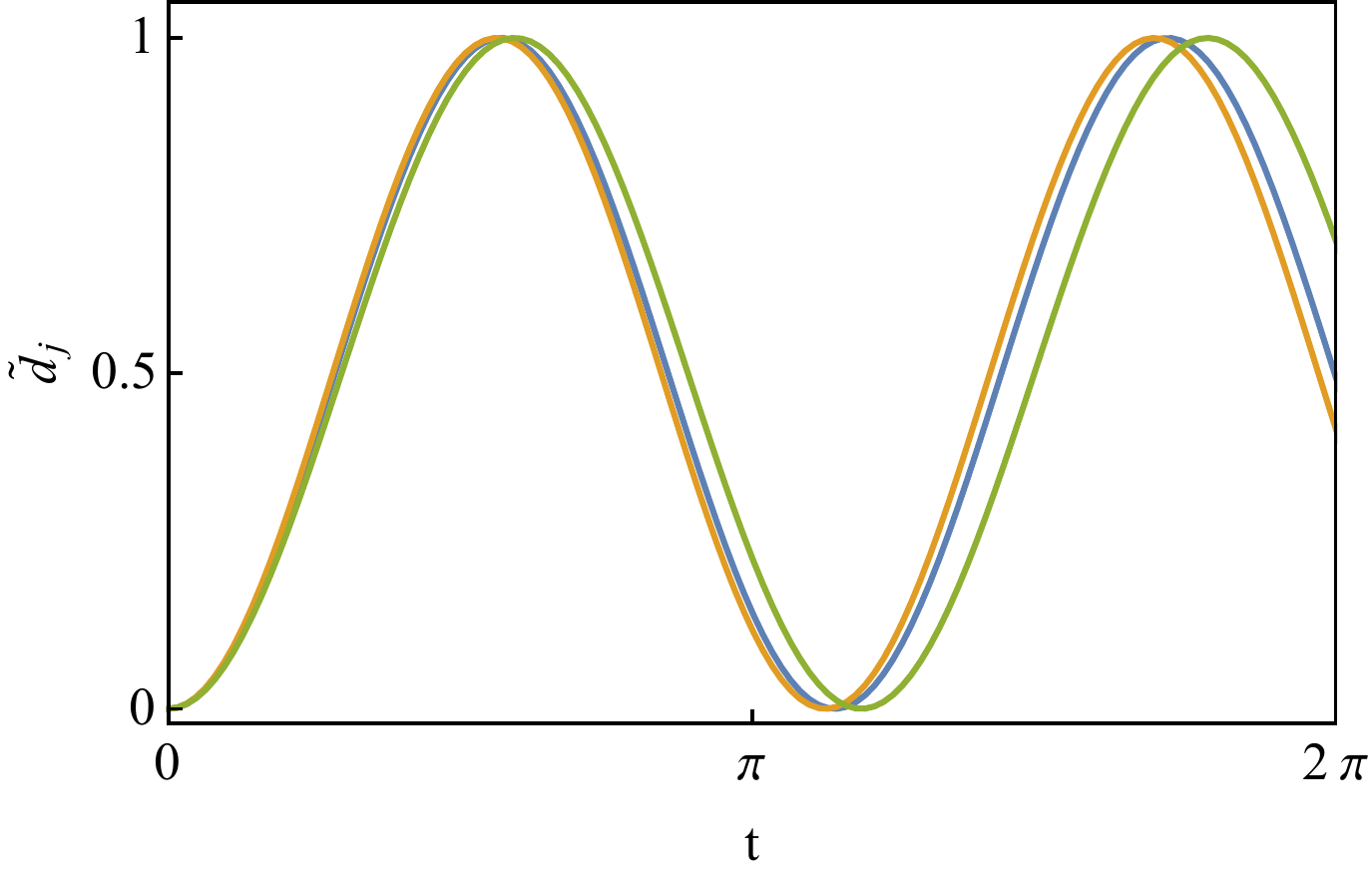}
\end{minipage}
}
\\
\subfloat[][]{
\begin{minipage}[t]{0.5\textwidth}
\flushleft
\includegraphics[width=0.8\textwidth]{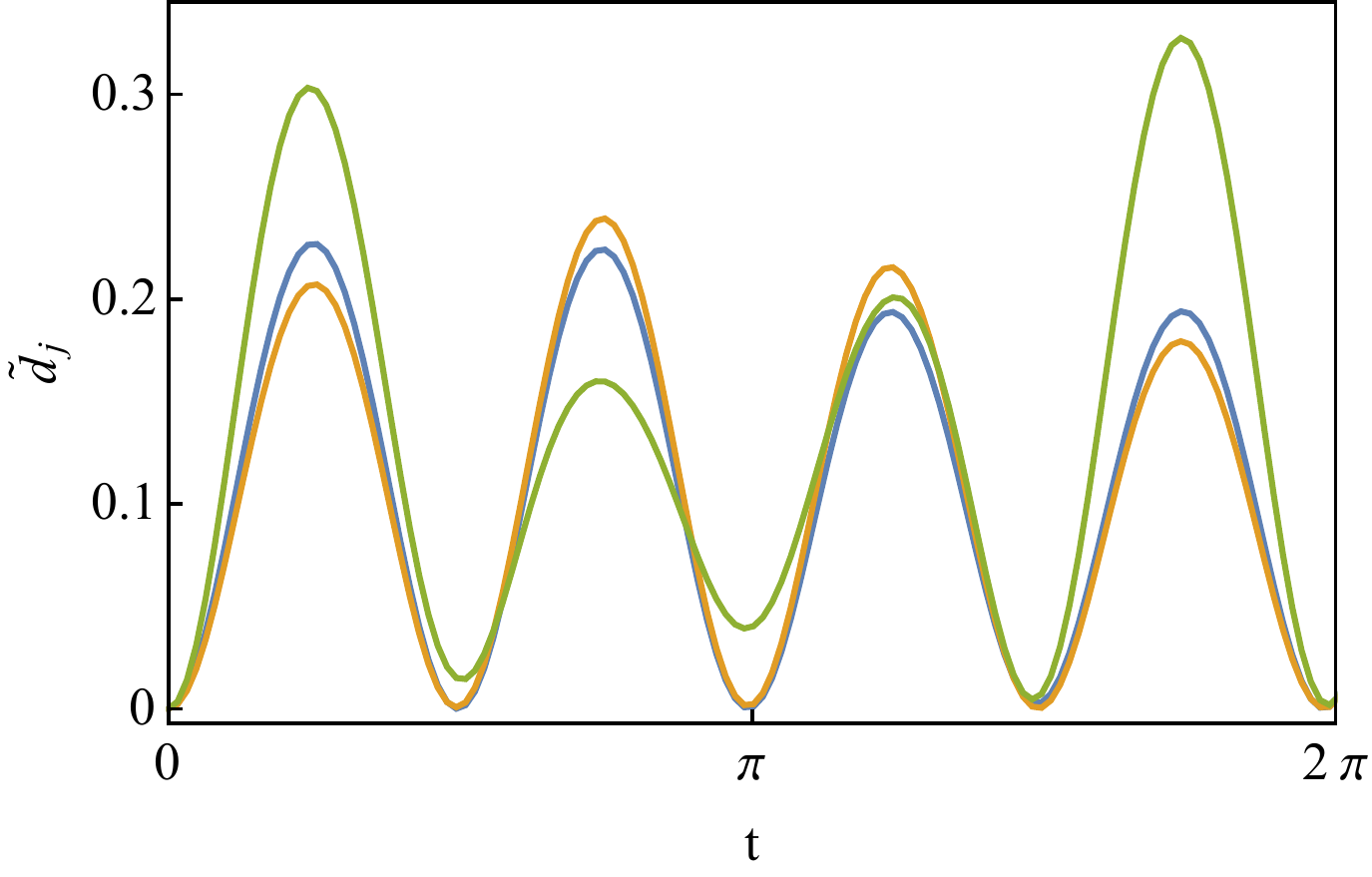}
\end{minipage}
}
\caption{The time evolution of the probability, $\tilde{d}_j=|c_{jj}(t)/c_{j0}(0)|^2$, to occupy the internal molecular state $b$ of an individual molecule in a molecular cluster ($N=60$) that starts at $t=0$ from the bright state $|B\rangle|V_0\rangle$ out of cavity (panel (a)) or from the lower $a$-polariton, Eq. \eqref{eq:inimj1}, inside a cavity, characterized by $\Omega_R=\sqrt{60}$ (panel (b)). The evolution is shown for three randomly chosen individual molecules. In these simulations the parameters $\lambda_j$ for different molecules are sampled from a uniform distribution in the range $1\pm0.2$. This random choice is reflected in the different periods of the internal Rabi oscillations seen out of the cavity. In contrast, the oscillations seen in the cavity are in phase, with a period that was found (not shown) to depend on $\Omega_R$, although with variable amplitudes.}
\label{fig:randoml}
\end{figure}

Figures \ref{fig:collective} and \ref{fig:randoml} demonstrate the effect of the cavity environment on the internal molecular dynamics following polariton excitation. The sums $r_0(t),d_0(t)$ and $d_1(t)$ (Eqs. \eqref{eq:r0t}-\eqref{eq:d1t}) are plotted against time in Fig. \ref{fig:collective}. Also shown are the out of cavity behaviors of $d_0(t)$, in Fig. \ref{fig:collective}a, and $d_1(t)$ ($=1-d_0(t)$ when out of cavity), in Fig. \ref{fig:collective}b. Out of the cavity, the internal $a\rightarrow b$ dynamics is characterized by Rabi oscillations with a frequency that reflects the internal coupling $\lambda$. In the cavity, with increasing $N$, the long time dynamics of the population of the subspace in which all molecules remain in their internal $a$ state (referred to below as the molecular all-$a$ manifold), as embodied in the sum $d_0(t)$, become overall slower (Fig. \ref{fig:collective}a), however it shows fast local oscillations whose period is determined by the collective Rabi splitting $\Omega_R$. Surprisingly, unlike the out of cavity dynamics, in the cavity the time evolution of $d_0(t)$ (Fig. \ref{fig:collective}a) and $d_1(t)$ (Fig. \ref{fig:collective}b) are not correlated. Instead, in the cavity, $d_1(t)$ shows oscillations of considerable amplitude that correlate with the evolution of the cavity mode population (represented by $r_0(t)$ in Fig. \ref{fig:collective}c). As shown in Fig. \ref{fig:collective}c, this apparent population exchange between the cavity mode and molecular states $b$ depends on the collective Rabi splitting $\Omega_R$. Another view of the system given by the time evolution of $r_1(t)$ (cavity mode is excited and the molecular system is in state $|V_j\rangle,j>0$) is seen in Fig. S3 in SI.

In Figure \ref{fig:randoml}. we show another interesting difference between the time evolution in and out of the cavity. Here, a molecular system $(N=60)$ for which the internal coupling parameters $\lambda$ are chosen randomly about some average value (here taken $1$), evolves out of the cavity following excitation of the bright state, and in the cavity following excitation of the lower $a$-polariton. The probabilities, $\tilde{d}_j(t)$, for individual excited molecules $j$ to make the transition to internal states $b_j$ are shown as functions of time. Out of cavity the molecules evolve independently, and each molecule displays an internal dynamics characterized by its own Rabi oscillation between internal states $a$ and $b$. In contrast, in the cavity all molecules appear to synchronize and oscillate in phase, albeit with varying amplitudes that reflect their relative weight in the molecular bright mode. A similar behavior is seen when the random couplings $\lambda_j$ are augmented by random phase terms (see Fig. S4 in the SI where a larger molecular system $(N=120)$ is used). This apparent synchronization reflects the fact that collective coupling of the $a$ states to the cavity mode opens an energy gap between the $a$ and $b$ states and the oscillation frequency between these states is now dominated by this gap.

\begin{figure}[t]
\includegraphics[width=0.8\textwidth]{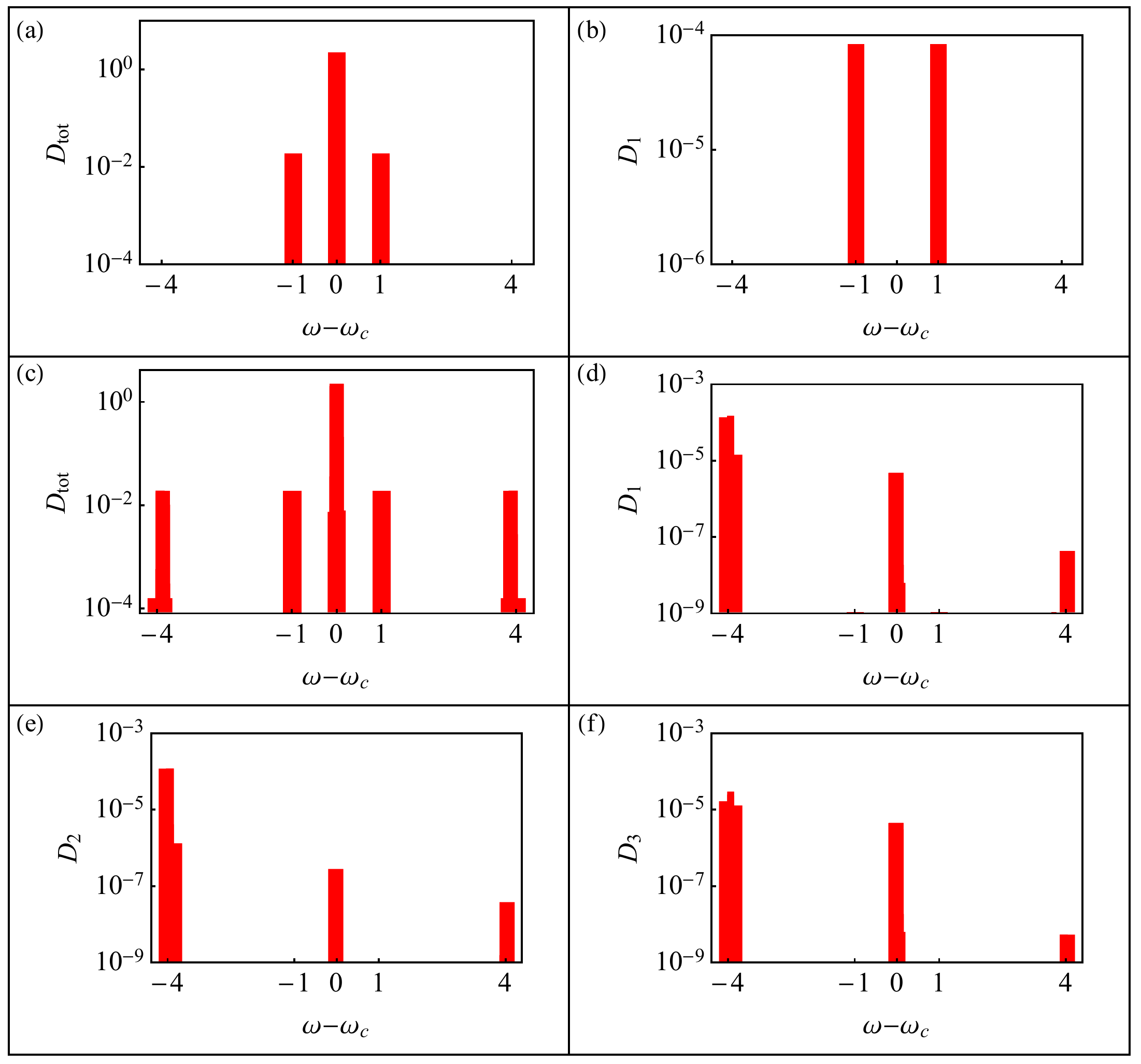}
\caption{Panel (a) and (b): (weighted) density of states outside the cavity. Panel (c-f): (weighted) density of states inside the cavity. The size of the molecular cluster is $N=120$. Parameters are $\Omega_R=\sqrt{60},g=1/\sqrt{2},\omega_c=\omega_{xg}(=0)$. The internal coupling is $\lambda=1$ for both inside and outside the cavity. Resolution is $\delta\omega=0.4$.}
\label{fig:dosU}
\end{figure}

Further insights into these interesting cavity effects can be obtained from different views of the density of states, shown in Fig. \ref{fig:dosU}. These densities were generated for a system with $N=120$ molecules (other parameters are $g=1/\sqrt{2},\Omega_R=\sqrt{60}$ and $\lambda=1$). Figure \ref{fig:dosU}a and Fig. \ref{fig:dosU}b respectively show, for the molecular system out of cavity, the total density of states, $D_{tot}(\omega)$ and the density of states associated with the $a$-bright state, $D_1(\omega)$. In contrast to the total density of states that is dominated by the dark states about $\omega=\omega_{xg}$, the weighted density $D_1(\omega)$ (panel (b)) shows only the internal splitting $2\lambda$ that characterizes states in which one molecule is excited. Inside the cavity (panels (c)-(f)), the total density of states is again dominated by the dark modes. However, the weighted densities, Eq. \eqref{eq:wDOS}, calculated for $\Psi(t=0)$ given by the lower $a$-polariton, show the largest contribution near the energy of this initial polariton at $\omega-\omega_c=-\sqrt{60}/2$, while also showing a significant although much smaller peak near $\omega=\omega_c$ and even a small signal near the upper polariton energy. Comparing panels (5d), (5e) and (5f) we see that the density peak near $\omega-\omega_c=-\sqrt{60}/2$ is dominated by $a$-type states while states near $\omega\sim0$ are of the b type. Importantly, while outside the cavity the initial bright state contains equal contributions of  eigenstates of a and b character, inside the cavity the initially populated $a$-polariton has only a small weight among b-type states and therefore shows only a small evolution into this part of the system state space. Similar (weighted) DOS spectra for a system with a larger Rabi splitting ($\Omega_R=\sqrt{120}$), displayed in Fig. S5 in the SI, show similar trends.

\begin{figure}[!t]
\subfloat[][]{
\begin{minipage}[t]{0.5\textwidth}
\flushleft
\includegraphics[width=0.8\textwidth]{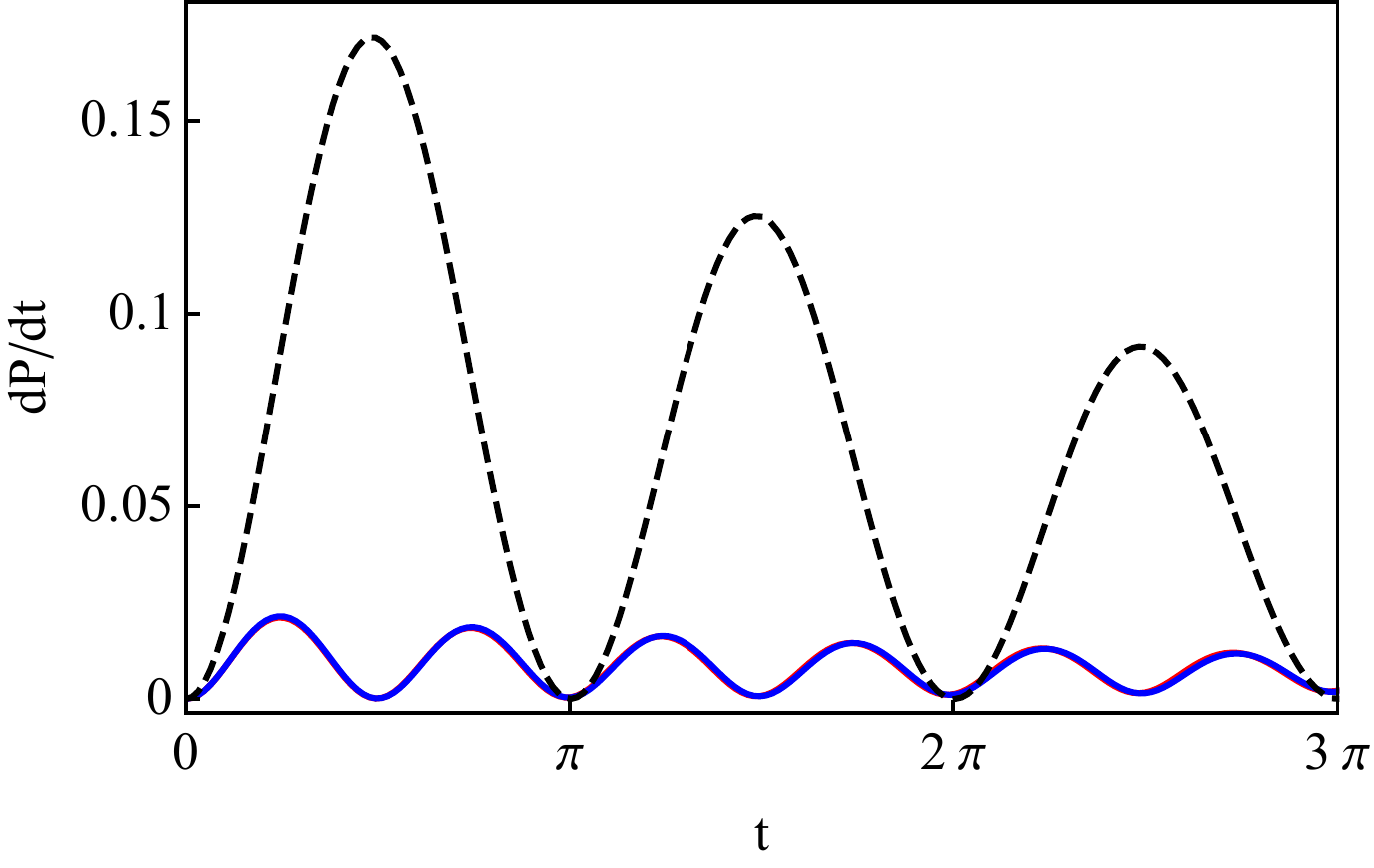}
\end{minipage}
}
\\
\subfloat[][]{
\begin{minipage}[t]{0.5\textwidth}
\flushleft
\includegraphics[width=0.8\textwidth]{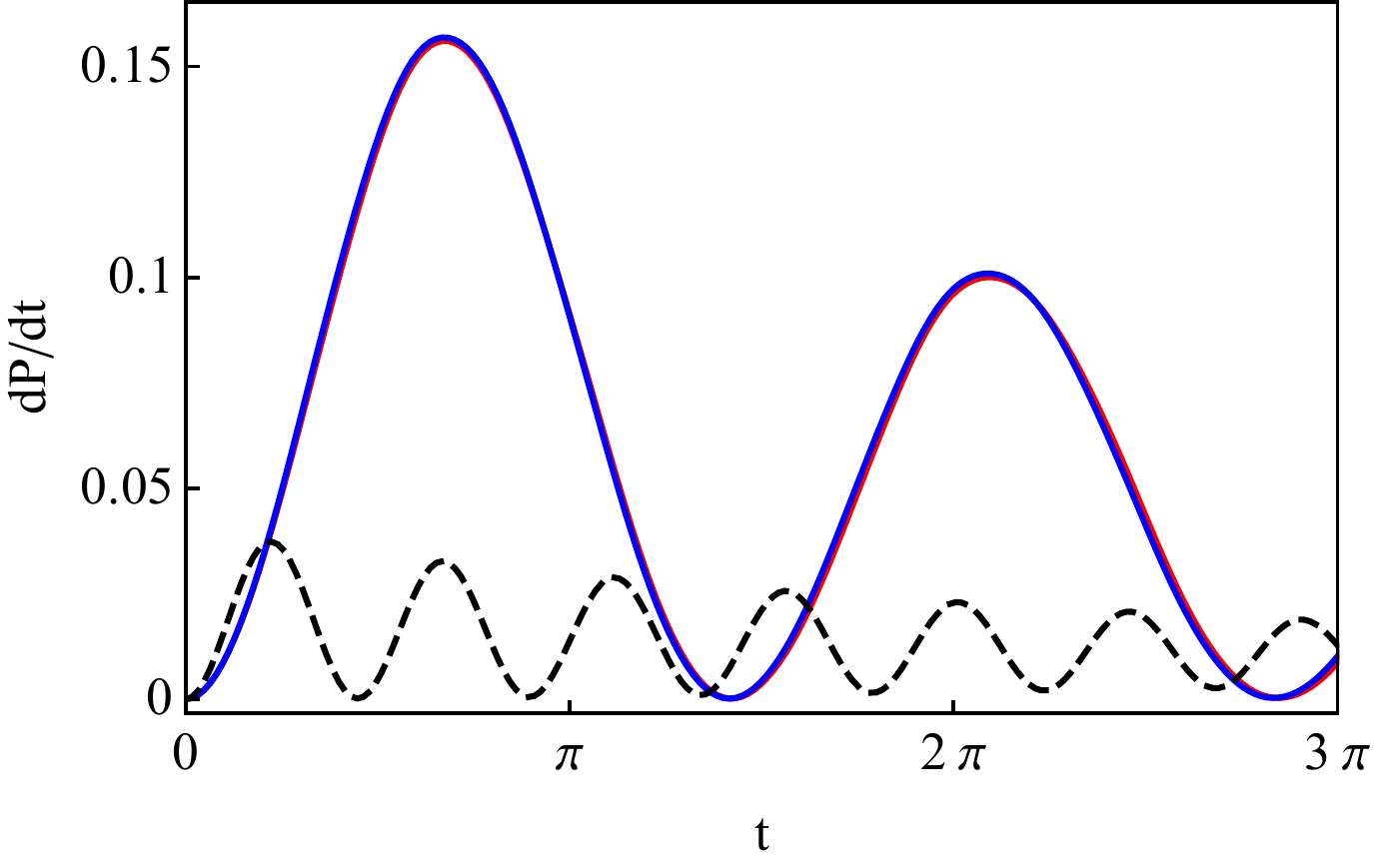}
\end{minipage}
}
\\
\subfloat[][]{
\begin{minipage}[t]{0.5\textwidth}
\flushleft
\includegraphics[width=0.8\textwidth]{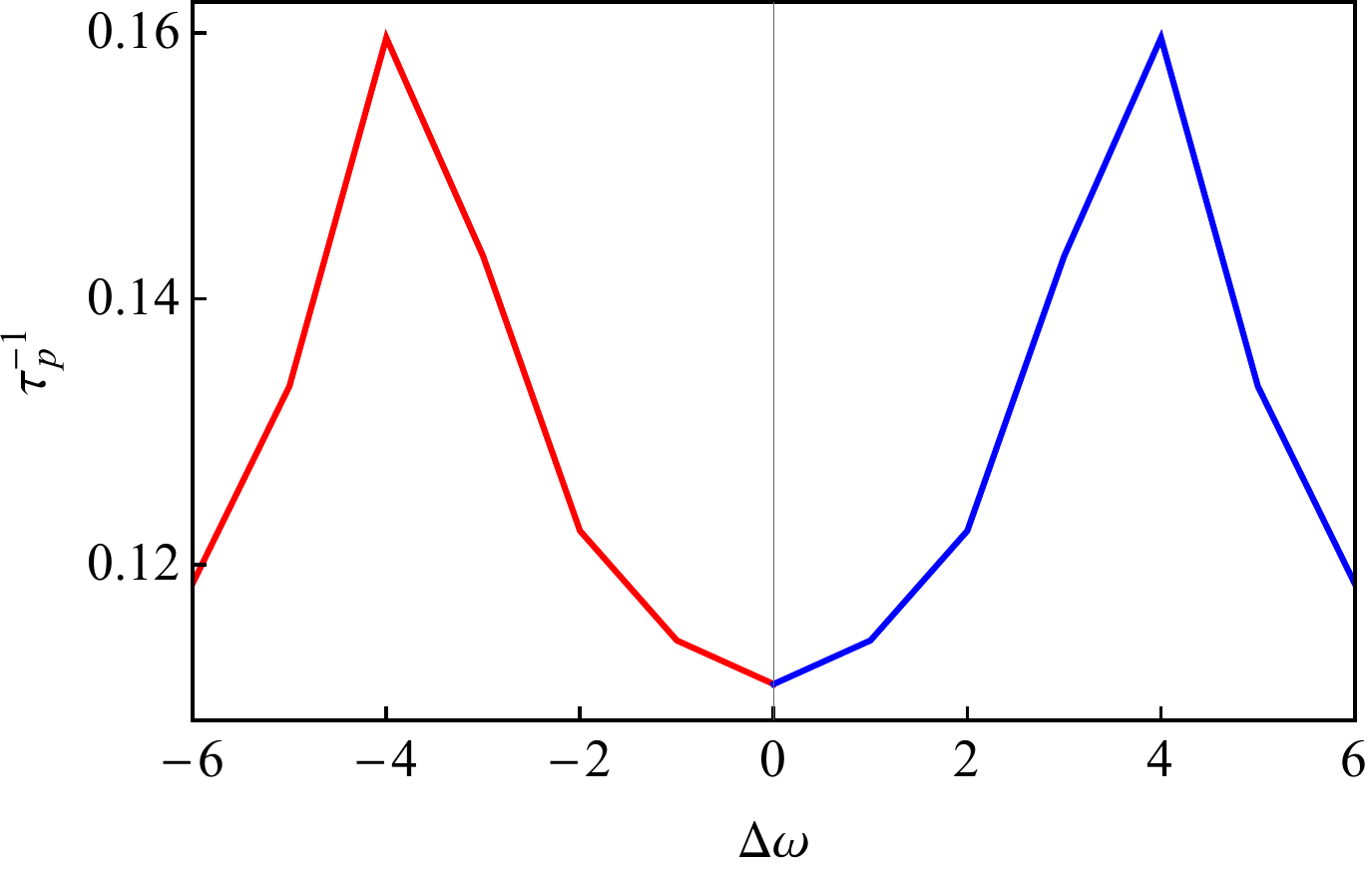}
\end{minipage}
}
\caption{The instantaneous rate of product formation and the inverse product formation time, represented by the damping associated with states $|j,j\rangle$. (a) The instantaneous rate of product formation defined in Eq. \eqref{eq:dPt} for the case where the system is inside the cavity evolving from the lower $a$-polariton, Eq. \eqref{eq:initial}, with $ \Omega_R=\sqrt{60}$ (red solid line, $N=60$; blue solid line, $N=120$) or outside the cavity evolving from the bright state $|B\rangle|V_0\rangle$ (black dashed line - this result is independent of $N$). In all cases, the energy gap between two inner states is $\Delta\omega=\epsilon_b-\epsilon_a=0$ and the damping coefficient is $\eta_b=0.2$. (b). Same setting as in (a) except for the energy gap $\Delta \omega=-4$. Note that in panels (a) and (b), blue and red lines overlap. (c) The product formation rate $\tau_p^{-1}$, Eq. \eqref{eq:prate}, as a function of the energy gap $\Delta \omega$, between the molecular states $a$ and $b$. The initial state is either the lower $a$-polariton (red), or the upper $a$-polariton (blue). The system size is $N=120$, $\Omega_R=\sqrt{60}$ and the damping coefficient is $\eta_b=0.2$. Other parameters are the same as as Fig. \ref{fig:dosU}. }
\label{fig:damp}
\end{figure}

Another consequence of the cavity effect on the $a\rightarrow b$ evolution following excitation of the lower $a$-polariton is shown in Fig. \ref{fig:damp}. Here, the evolution of b states is augmented by damping terms, representing a process in which the molecule in state $b$ can transport into some product at a rate $\eta_b$. The rate of product formation, Eq. \eqref{eq:dPt}, is plotted in Fig. \ref{fig:damp}a, for a molecular system inside and outside the cavity, where the energy spacing between the $a$ and b molecular states is taken $\Delta \omega=0$. Outside the cavity, the molecules evolve independently, and the ($N$ independent) product formation rate following the excitation of the single $a$-bright exciton state reflects the periodic variations in populations of molecular states $b$ with period $\lambda$. Inside the cavity, this process becomes significantly slower and the rate is seen to decrease as the number of molecules increases at constant collective Rabi frequency. This behavior appears to reflect the polaron decoupling effect based on the cavity-Born-Oppenheimer picture that would imply that the effective internal state coupling $\lambda$ is reduced, $\lambda\rightarrow\lambda/N$. However, Fig. \ref{fig:damp}b offers a more conventional interpretation while showing a more complex behavior: when $\Delta\omega=E_b-E_a=-\Omega_R/2$, namely when state $b$ is placed in proximity with the lower polariton, the rate inside the cavity is enhanced while that out of the cavity becomes slower. The similarity of the rates shown in Fig. \ref{fig:damp}a out of cavity and in Fig. \ref{fig:damp}b in the cavity suggests that the main effect seen is not renormalization of $\lambda$ but proximity of energy levels. Another view of the same effect is seen in Fig. \ref{fig:damp}c, where another measure for the product formation rate, the inverse average product formation time, Eq. \eqref{eq:prate}, is plotted as a function of $\Delta\omega$. The maximal rate is obtained when state $b$, the doorway to product formation in our model, is close in energy to the upper and lower polariton. For the same system size $N$ but with $\Omega_R=\sqrt{120}$ $(g=1)$, we show the product formation rate in Fig. S6 in the SI.

The results displayed in Figs. \ref{fig:collective}-\ref{fig:damp} (as well as the additional Figures S3-S8 shown in SI) lead us to conclude that the polaron decoupling picture, by which strong coupling between molecular electronic and cavity mode transitions leads to nuclear dynamics in the bright single-exciton that is dominated by the ground state potential surfaces, is not seen in the dynamics of our model. We recall that as in the HTC model, our model is characterized by internal slow internal dynamics of individual molecules, except that the harmonic oscillator (representing nuclear motion) in the HTC model is replaced by a 2-level $(a,b)$ system. The vibronic coupling (manifested as a difference between the harmonic potential surfaces in the upper and lower molecular electronic states) is replaced by the difference between the internal interlevel coupling, assumed to be zero in the ground and finite $(\lambda)$ in the upper electronic state. The Condon approximation that implies that electronic excitations take place at a fixed nuclear coordinate is replaced by the assumption that electronic transitions occur at constant internal molecular state so that under the model assumption that the stable molecular state in the ground electronic state is $a$, a short time pulse will excite the $a$-polariton in which all molecules are in the $a$ state. Considerations similar to those that lead to the polaron decoupling picture would predict that internal dynamics following such excitation that transforms $a$ to $b$ states will be characterized by an effective internal coupling $\lambda/N$ where $N$ is the number of molecules involved in the bright state that makes the polariton. While the slow (relative to $\lambda^{-1}$) overall evolution of the initially formed $a$ population seen in Fig. \ref{fig:collective}a appears to support this picture, the fast oscillations that correlate with Rabi frequency $\Omega_R$ suggest that the dominant effect underlying the observed dynamics is the gap of this order that opens, because of the collective coupling to the cavity modes, between $a$ states that make the excited polariton, and $b$ states that do not collectively couple to the cavity mode. The results displayed in Fig. \ref{fig:randoml} (as explained in the discussion of this Figure) and the weighted density of states studied in Figure \ref{fig:dosU}, further support this conclusion.

Underlying the disparity between our observations and the polaron decoupling picture is the fact that because of timescale separation between the fast electronic and cavity dynamics and the internal dynamics of individual molecules, and in the spirit of the Born approximation and its analog assumed by our present model, the initially excited polariton reflects the instantaneous configuration of the slow molecular motion - the instantaneous nuclear configuration in the HTC model or $a$-polariton in our model. In the space of the slow coordinate the collective coupling to the cavity is local (in our case – in the $a$ subspace since we assumed that the molecular ground state is of type $a$), a collective analog of the observation made in single molecule calculations, see e.g. Ref. \cite{Kowalewski2016,Mukamel2016}. The subsequent $a\leftrightarrow b$ dynamics is then dominated by the fact that the collective coupling to the cavity mode opens a gap between $a$ and $b$ states (or closes a gap that exists outside the cavity). This is clearly seen in the study of the model in which a reactive channel is opened out of states $b$ as seen in Fig. \ref{fig:damp} and explained in the accompanying discussion. Indeed, the experimental results of Refs. \cite{Avramenko2020,Eizner2019,Esteso2021} appear to reflect this behavior.

One observation in Figures \ref{fig:collective}(b-c) appears to lie outside this picture: it is not immediately clear why, following the creation of the a-polariton, the dynamics of populating the $b$ internal states appear to be correlated with the cavity mode dynamics rather than with the evolution of the population on the $a$ molecular subspace. It can be shown however that this behavior is associated with the local (a-type) nature of the initial state and can be explained using a 3-state $(A,B,C)$ model in which $A$ represents the $a$-bright state of the molecular system that forms the initially excited polariton, $B$ represents the $b$-type state that evolves out of it due to the coupling $\lambda$ and $C$ is the amplitude of the excited cavity mode. A minimal dynamic model described by (see Sec. IV in SI)

\begin{align}
\frac{dA}{dt}&=-i\Omega_RC-i\lambda B,\notag\\
\frac{dC}{dt}&=-i\Omega_RA,\notag\\
\frac{dB}{dt}&=-i\lambda A,
\end{align}
(in our case, for the lower $a$ polariton $A(t=0)=-C(t=0)=-2^{-1/2}; B(t=0)=0$) shows that when $\Omega_R\gg\lambda$, the evolution of the $b$-subspace is correlated mainly (in this minimal model) with that of the cavity mode.

\begin{figure}[t]
\subfloat[][]{
\begin{minipage}[t]{0.5\textwidth}
\flushleft
\includegraphics[width=0.8\textwidth]{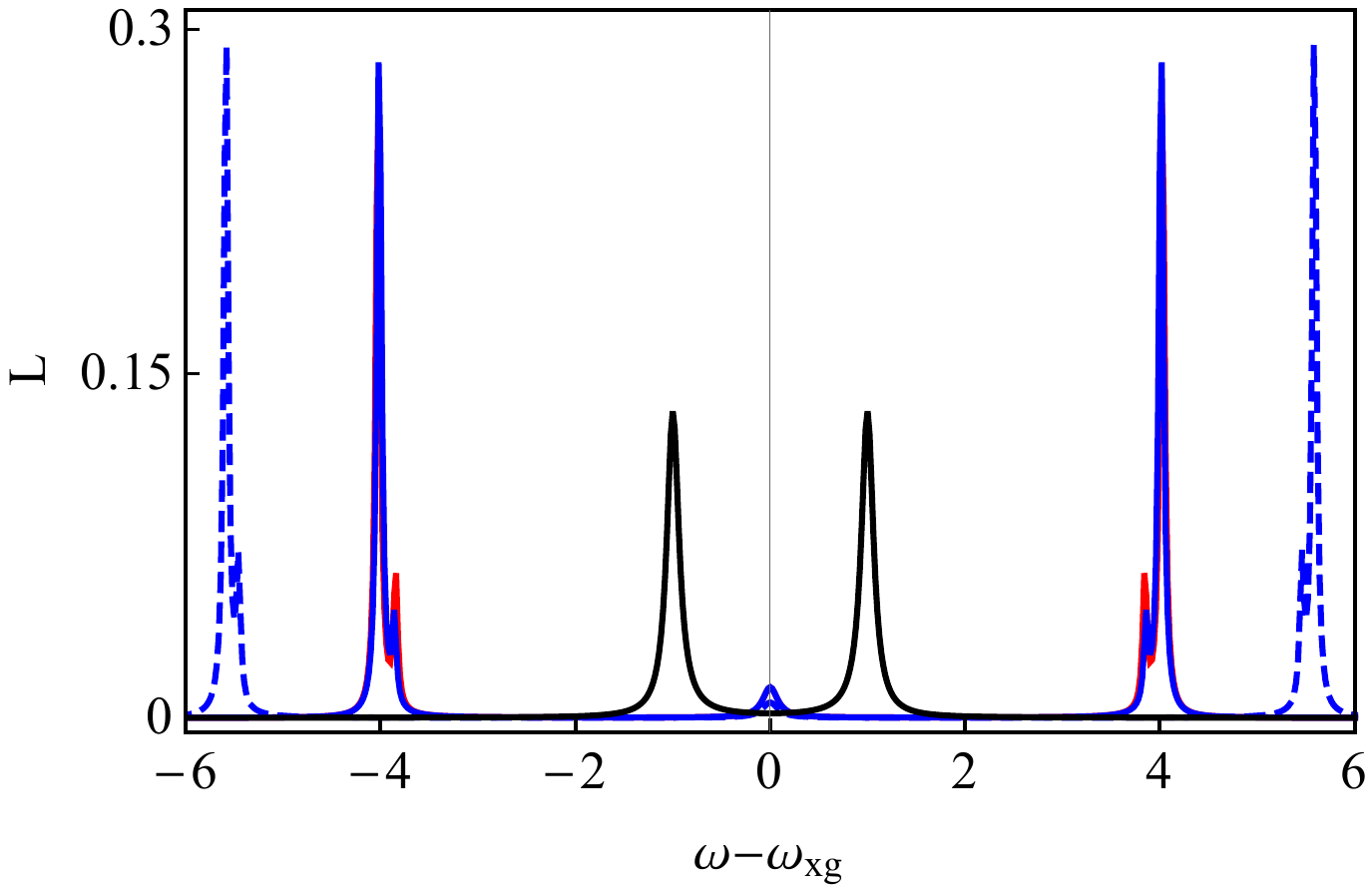}
\end{minipage}
}
\\
\subfloat[][]{
\begin{minipage}[t]{0.5\textwidth}
\flushleft
\includegraphics[width=0.8\textwidth]{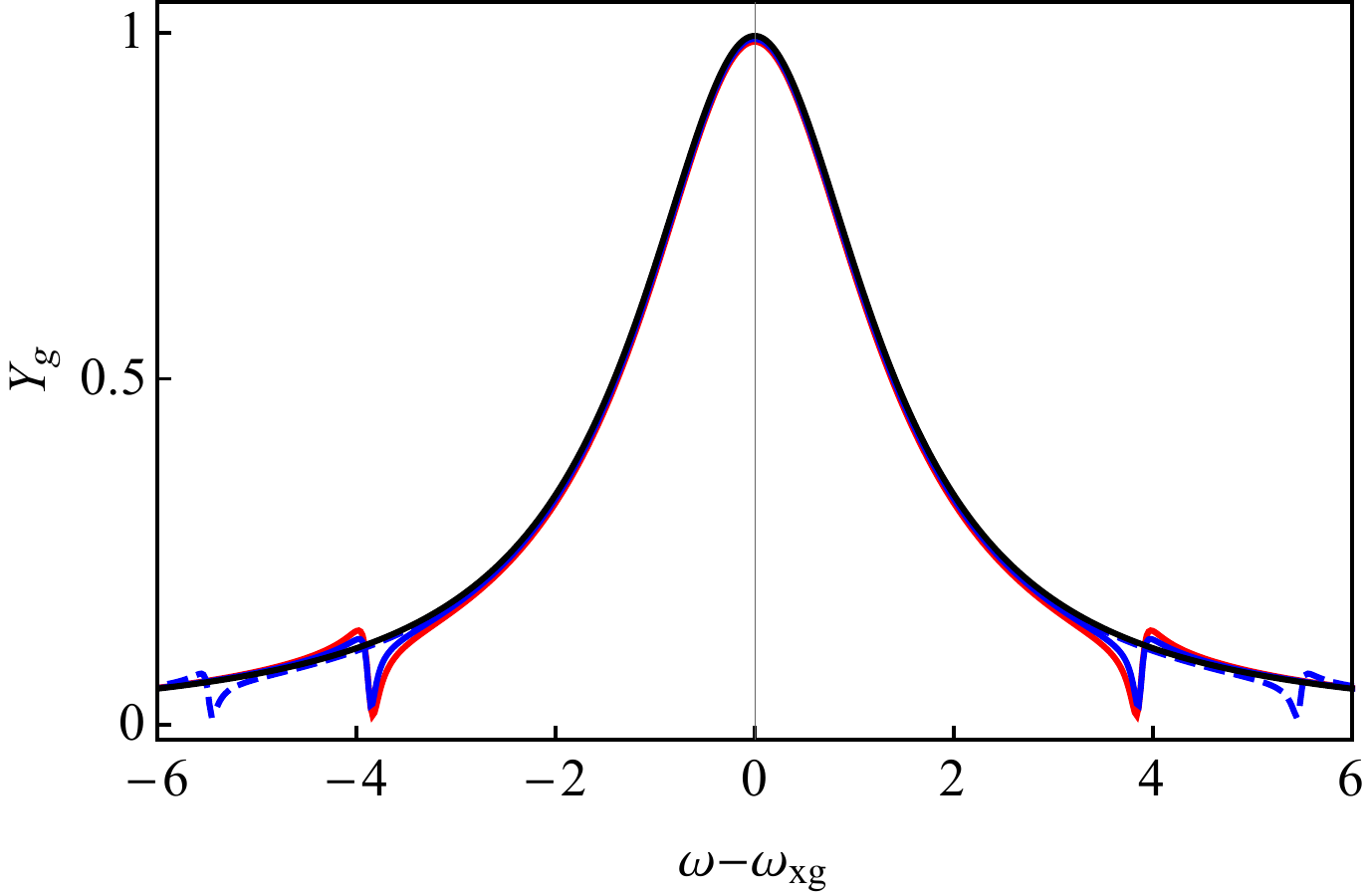}
\end{minipage}
}
\caption{(a) The absorption lineshape (in arbitrary units) and (b) the product yield, plotted against the driving frequency measured from the single molecule transition frequency, when the system is inside (red lines for $N=60$ and $\Omega_R=\sqrt{60}$, full blue lines for $N=120, \Omega_R=\sqrt{60}$ and dashed blue lines for $N=120, \Omega_R=\sqrt{120}$ ($g=1,1/\sqrt{2}$ and $1$, respectively)) or outside the cavity (black lines). The energy gap between two inner states $a$ and $b$ is $\Delta\omega=0$ and the damping coefficients are $\eta_a=0.1,\eta_b=0.2$. Other parameters are the same as Fig. \ref{fig:damp}. Note that in (b), the different lines overlap in most of the shown selected region.}
\label{fig:CW}
\end{figure}

Finally, in Fig. \ref{fig:CW} we show results of a CW calculation based on the formalism described in Eqs. (\ref{eq:G}-\ref{eq:yield}). Here the molecular system is driven from the ground state $|G\rangle|V_0\rangle$, see Eq. \eqref{eq:GV0} by an external incident photon field of frequency $\omega$. This driving state is assumed to couple to the $a$-bright state, Eq. \eqref{eq:brightvibro}, of the molecular system when it is outside the cavity, and to $|X_0\rangle|V_0\rangle$ that the cavity mode is excited with all molecules unexcited and in state $a$, when it is inside the cavity. These states are not eigenstates of the full system Hamiltonian because of the internal molecular coupling $\lambda$, but provide doorways towards populating the excited molecular states. Excited molecular electronic states with inner states $a$ and $b$ are further assumed to undergo relaxation into external channels, represented by damping rates $\eta_a$ and $\eta_b$. To facilitate comparison between in and out of cavity spectra and focus on collective molecular effects promoted by the cavity environment we have suppressed the loss rate $\eta_0$ associated with the cavity mode. (Results of calculations that include this loss are shown in Fig. S7 in the SI). At the steady state, the total combined flux into $a$ and $b$ relaxation channels can be identified as the absorption lineshape, Eq. \eqref{eq:lineshape}, while the quantum yield, Eq. \eqref{eq:yield}, is the ratio between the flux going out of states $b$ and the total absorption. Such representative spectra are shown in Figs. \ref{fig:CW}a and \ref{fig:CW}b, respectively ((see also Fig. S7 in the SI for the case where the loss is taken to be dominated by the cavity mode)). The absorption spectrum shows the expected structures: out of cavity (where the $g\rightarrow0$ limit of the Tavis-Cummings model shows a molecular peak about $\omega=\omega_{xg}$), our model shows the split peak associated with the internal molecule coupling $\lambda$. Inside the cavity, here characterized by $\Omega_R=\sqrt{60}$, the upper and lower $a$-polariton lines dominate the spectrum, while a weak signal from the $b$-state manifold is seen near $\omega=0$. It is interesting to note that the effect of the intramolecular coupling $\lambda$, which is seen prominently in the out of cavity spectrum practically disappears inside the cavity when $\Omega_R\gg \lambda$, where most of the all-$a$ states are pushed to the polariton bands. This observation is consistent with the density of state analysis, as seen by comparing, for example, Fig. \ref{fig:dosU}b and \ref{fig:dosU}d.

Turning to the yield spectrum, Fig. \ref{fig:CW}b, two observations are notable: First, the sharp structures that characterize the absorption spectra (and in the absolute product formation) are not seen in the yield, which is their ratio. This is consistent with the well-known behavior of models characterized by simple Lorentzian lineshapes. The Lorentzian width $\Gamma=\sum_j\Gamma_j$ is the sum of relaxation rates $\Gamma_j$ associated with different channels, and the yield of a given channel, $Y_j=\Gamma_j/\Gamma$ is energy independent. In the present model, the broad peak about $\omega=0$ is, remarkably, very similar in and out of the cavity.

Secondly, again remarkable, are the dips observed in the quantum yield about the frequencies of the upper and lower polaritons, that mark the main difference between the in- and out- of cavity situations. The appearance of these dips near the most prominent polaritonic absorption peaks is a significant cavity effect. While it is tempting to rationalize this phenomenon as a manifestation of the analog of polaron decoupling (that would lead to a smaller effective coupling $\lambda$ toward the product channel), we note that the absorption at the polaritonic peaks is dominated by the excitation of the $a$-polaritons and relaxation via $\eta_a$ processes, while the product channel is promoted by populating the $b$-states (with energies near $\omega=0$) following by their $\eta_b$ relaxations. This combination of state distribution and their decay rates may also lead to the observed dips. The quantum yield dip near the polariton resonance is consistent with a recent calculation \cite{Sukharev2022} in which the basis truncation is avoided while other approximations - classical representation of the radiation field and mean field treatment of the molecular subsystems, are invoked. A word of caution is in order, though. Recall that the smallness of $d_2$, Eq. \eqref{eq:d2t}, was suggested as a plausibility criterion for the validity of our basis truncation approximation. We have found that for above choice of parameters, this criterion is well satisfied for all the driving frequency $(\omega)$ range except near the polariton frequency where the dip appears (c.f. Fig. S8 in SI). For $\omega$ in this neighbourhood, a significant long-time $d_2$ population is built in the system appears to dominate the loss flux, casting doubt on this interesting dip observation. Further studies will be needed to establish the significance of this observation.

\section{Conclusion}
\label{sec.5}
We have used a simplified model akin to the Holstein-Tavis-Cummings model, in which the harmonic oscillator representing the molecular nuclear motion that couples to the molecular electronic transition is replaced by a 2-level system. The slow ("nuclear") dynamics in our model is represented by a parameter $(\lambda)$ that plays the role of vibronic coupling (being zero and finite in the lower and upper electronic state, respectively). It also determines the timescale for the internal molecular motion that follows an electronic excitation. Another model parameter $\Delta\omega$ has been used to describe situations where the internal coupling $\lambda$ connects between states of different energies, as encountered for example when the excitation induces transitions between different electronic states.

Our simplified model, together with judicious basis truncation has made it possible for us to investigate the short time collective dynamics of a 1-exciton state comprising $N$ molecules and one cavity mode and to clarify the origin and nature of collective response in such systems. While not representing a real molecular system, we could use this model to elucidate the consequences of the interplay between the collective nature of the system electronic response and the local nature of the (slow, in analogy to molecular nuclear dynamics) internal motion of individual excited molecules. Our observations are not compatible with the polaron decoupling picture that was suggested as a possible origin for the effect of cavity environments on charge transfer processes. Rather, they are consistent with dynamics dominated by the polaritonic shift between excited states that are accessible from the system ground state and those that are not.

We have also investigated the response of our model system to CW driving in the linear response regime. The appearance of a dip in the yield of product formation out of the $b$ states at the polariton energy could be interpreted as a manifestation of the polaron decoupling mechanism, but in light of our other observations we again suggest the aforementioned polaritonic as the origin. We note that this observation is not conclusive because of the limitations of our numerical procedure and should be subject to further scrutiny.

The calculations reported in this work correspond to a zero temperature system that does not interact with its environment. Dephasing thermal relaxation should have a profound effect on the observed collective response \cite{Zeyu2022}. Semiclassical mean-field approximations that have proven useful in coupled plasmon-exciton systems \cite{Sukharev2017} will be explored to this end in future work. 

\section*{Supplementary Material}
See supplementary material for discussions of the time evolution of a single excited molecular state in the Tavis-Cummings system, more figures in addition to those shown in Sec. IV and a toy model of a 3-level system.

\section*{Acknowledgements}
This material is based upon work supported by the U.S. National Science Foundation under Grant CHE1953701. We thank Joe Subotnik and Maxim Sukharev for many useful discussions.

\section*{Author Declarations}
\subsection*{Conflict of interest}
The authors have no conflicts to disclose. 

\section*{Data availability}
The data that support the findings of this study are available within the article [and its supplementary material].

\bibliography{reference}

\end{document}